# Revealing electron-electron interactions in graphene at room temperature with the quantum twisting microscope

M. Lee[1,2]\*, I. Das[1,2]\*, J. Herzog-Arbeitman[3]\*, J. Papp[1,2], J. Li[1,2], M. Daschner[1,2], Z. Zhou[4], M. Bhatt[1,2], M. Currle[1,2], J. Yu[5], Yi Jiang[6], M. Becherer[7], R. Mittermeier[7], P. Altpeter[1,2], C. Obermayer[1,2], H. Lorenz[1,2], G. Chavez[1,2], B. T. Le[1,2], J. Williams[1,2], K. Watanabe[8], T. Taniguchi[8], B. Andrei Bernevig[3,6,9] and D. K. Efetov[1,2]†

1. Fakultät für Physik, Ludwig-Maximilians-Universität, München, Germany
2. Munich Center for Quantum Science and Technology (MCQST), München, Germany
3. Department of Physics, Princeton University, Princeton, New Jersey 08544, USA
4. School of Physics, Peking University, Beijing 100871, China
5. Department of Physics, University of Florida, Gainesville, FL, USA
6. IKERBASQUE, Basque Foundation for Science, Bilbao, Spain
7. School of Computation Information and Technology, Technical University of Munich, Germany
8. National Institute of Material Sciences, Tsukuba, Japan
9. Donostia International Physics Center, Donostia-San Sebastian, Spain

\*These authors contributed equally to this work; † E-mail: dmitri.efetov@lmu.de;

**The Quantum Twisting Microscope (QTM)[1-3] is a groundbreaking instrument that enables energy- and momentum-resolved measurements of quantum phases via tunneling spectroscopy across twistable van der Waals heterostructures. In this work, we significantly enhance the QTM's resolution and extend its measurement capabilities to higher energies and twist angles by incorporating hexagonal boron nitride (hBN) as a tunneling dielectric. This advancement unveils previously inaccessible signatures of the dispersion in the tunneling between two monolayer graphene (MLG) sheets—features consistent with a logarithmic correction to the linear Dirac dispersion arising from electron-electron (e-e) interactions with a fine-structure constant of $\alpha \approx 0.32 \pm 0.01$[4-18]. Remarkably, we find that this effect, for the first time, can be resolved even at room temperature, where these corrections are extremely faint. Our results underscore the exceptional resolution of the QTM, which, through interferometric interlayer tunneling, can amplify even subtle modifications to the electronic band structure of two-dimensional materials. Our findings reveal that strong e-e interactions persist even in symmetric, non-ordered graphene states and emphasize the QTM's unique ability to probe spectral functions and their excitations of strongly correlated ground states across a broad range of twisted and untwisted systems.**

## Main

The quantum twisting microscope (QTM)[1-3] is a recently developed technique that allows for dynamical in-situ twisting of individual 2D layers with respect to one another. It has been recently used to visualize electronic band structures[1], probe phonon dispersions[2] and measure the sub-moirés potential landscape in graphene-based systems[3]. By introducing insulating spacers between two semi-metallic 2D sheets, it was shown that electrons can coherently and elastically tunnel across the layers when their Fermi surfaces overlap and energy and momentum conserving transitions are matched. Through application of a bias voltage $V_b$ and varying the relative twist angle $\theta$ between the layers, it is hence possible to control the energy $E$ and momentum $k$ overlap of the Fermi surfaces, and to reconstruct the band structures of the individual layers. This capability of the QTM invoked comparisons to traditional band structure measurement techniques such as ARPES[5,6,10,11,19] and SdH oscillations[4].

Compared to these techniques, however, the QTM possesses several new attributes and capabilities. The QTM combines the high energy resolution of STM and SdH with the spectral momentum information of ARPES — but with additional gate tunability. As it relies on coherent tunneling processes across large, ultra-clean 2D interfaces, its sensitivity is only limited by spectral lifetime and temperature broadening, as well as the resolution of the measurement electronics, all of which can be highly optimized. The electrostatic tunability of the 2D layers allows for the mapping of a large energy window of $E \sim \pm\, 0.5$ eV above and below the Fermi energy (see supplementary information for the discussion on the gate-controlled electrostatics of the system), where the limit is primarily governed by the band gap of the dielectric and its breakdown strength. Since the QTM detects the Fermi surface overlap between the bands of two materials, it is highly sensitive to their dispersion relation across a broad energy and momentum range.

In this work, we perform angle-resolved tunneling measurements between two monolayer graphene (MLG) devices with a home-built room-temperature QTM. By using high-quality few-layer hexagonal boron nitride tunneling dielectrics, instead of the previously used WSe$_2$ layers[1], we are able to improve the resolution of the tunneling spectra and extend the range of the bias voltage by a factor of 3 up to $V_b = \pm\, 2.5$V. We can reach Fermi energies of $\mu \approx \pm 0.5$ eV within which clear deviations from linear Dirac dispersions are observable with excellent reproducibility. We observe several unanticipated features in our tunneling spectrum. Excluding more trivial effects through detailed and extensive modelling, we find the nonlinearity is the unique result of and an excellent match to velocity renormalization by electron-electron interactions, studied in the graphene literature only at low temperature using capacitance[12], SdH oscillations[4], scanning tunneling microscopy (STM)[20] and ARPES[5,6,10,11,19]. We assign the unique ability to directly resolve electron-electron interaction effects in momentum space, even at room temperature, to the new and powerful capabilities of the QTM[21-24].

**<u>QTM setup and device fabrication</u>**

The schematic illustrations of our home-built QTM setup and sample configuration are shown in Fig. 1a-b. Our tunneling junctions are formed between specially prepared pyramidal AFM tips that are coated with MLG and rotatable substrates that are co-laminated with hexagonal boron nitride (hBN)/few-layer Bernal graphene heterostructures. Here, the hBN flakes are chosen to be $\leq 4$ layers thick which act as high-quality tunneling barriers between the graphene sheets (see Supp. Info. for a detailed description of the setup and sample preparations).

As shown in Fig. 1b, the sample is voltage biased ($V_b$) in both D.C. and A.C. for simultaneous measurements of D.C. current ($I$) and differential conductance (d$I$/d$V$), and the resulting tunneling current is collected at the tip. The bias voltage is distributed between the classical charging energy and quantum capacitance effects, which set the Fermi levels of the graphene sheets. Voltage biasing the sample causes both the chemical potentials ($\mu_{T,S}$) of the tip (T) and sample (S), as well as the relative electrostatic potential ($\Phi$) to shift according to the electro-chemical balance $V_b = \Phi + \mu_S - \mu_T$ as illustrated in Fig. 1c. By rotating the sample with respect to the tip in real space, the respective Brillouin zones also rotate about the $\Gamma$ point introducing a momentum shift $k_\theta \sim K_D \theta$, where $K_D$ is the momentum of the Dirac point as illustrated. As is illustrated in Fig. 1d, this causes a separation of the $K$ ($K'$) points of the tip and sample in k-space.

## Twist-angle dependent tunneling spectroscopy

Fig. 1g shows the derivative of the differential tunneling conductance $d^2I/dV^2$ vs. $\theta$ and $V_b$ across the graphene (tip)/hBN/graphene (sample) (the corresponding $I/V$ and $dI/dV$ maps are shown in the SI), and Fig. 1f shows the corresponding line-cut of $d^2I/dV^2$ vs. $V_b$ for a fixed $\theta = 3.1°$. At biases $V_b$ below $< \pm 0.8$ V, we observe two distinct features where the tunneling current $I$ reaches a local maximum, which correspond to a large phase space of energy and momentum conserving tunneling transitions between the tip and sample MLG. These features are consistent with original QTM measurements[1], where band structure illustrations in Fig. 1h explain their origins. Here we focus on $d^2I/dV^2$ maps to resolve finer structures, where we identify a linear "x" shaped feature which marks the smallest twist-angle-dependent bias where the Fermi surfaces first overlap and allow momentum-conserving tunneling. This "onset" feature is lower intensity and more temperature-broadened than the prominent hourglass shaped feature at higher bias voltages. This sharp "nesting" feature corresponds to momentum-conserving tunneling processes being allowed across a large range of energies, corresponding to close nesting of the biased graphene bands.

The exact forms of the onset and nesting features in the $V_b$ vs. $\theta$ phase space are a direct consequence of the underlying dispersion relation of graphene. Assuming a perfectly linear dispersion relation of a Dirac band, the onset line occurs when the condition $V_b = \pm \hbar v_F K_D \theta$ is fulfilled[7], leading to a perfectly linear feature, where $\hbar$ is Planck's constant, $v_F$ is the effective Fermi velocity (to be distinguished from the bare velocity introduced momentarily). For a linear band dispersion, the nesting line instead corresponds to the relation[7] $\Phi(V_b) = \pm\hbar v_F K_D \theta$, and produces an hourglass feature in the $\theta - V_b$ plane due to quantum capacitance effects. Using the above formulas at low bias, it is possible to extract the effective Fermi velocity, these, and their generalizations, are derived in the theory appendix, which we find to be $v_F \sim 1.04 \times 10^6$ m/s. This value is in good agreement with earlier observations of the effective Fermi velocity model of linear Dirac bands in graphene[25,26] and the recent QTM measurements[1].

## Split nesting and non-linear onset conditions at high bias

At high bias $V_b > 0.8$ V however, Fig. 1g displays new features. Quite prominently, the nesting line splits into two branches. Less prominently but still noticeable, the onset feature deviates from linearity. These features are highly reproducible across 9 independent tunneling spectroscopy measurements arising from permutations of different tips, samples, and sample regions (see SI). This is shown in Fig. 2a, where the top two quadrants are data taken from two different real space locations for one tip/sample combination but are 60° apart in $k$-space. The two bottom quadrants are data taken on two different samples with the same tip. Overall, we find strong splitting features in all four quadrants, which all occur above the same $V_b > \pm 1$ V, and are symmetric between positive and negative values of $V_b$ and $\theta$. We further highlight the splitting behavior in a zoom-in image in Fig. 2b. It shows a characteristic line-shape, where the top branch splits and separates from the lower branch. At higher bias however, the top branch becomes stronger, while the bottom branch slowly fades out.

Both observed features are in direct contradiction to the linear Dirac model. For a linear dispersion the bands of the tip and sample MLG run in parallel over an extended line in $k$-space and hence allow for only one nesting condition. In general, the bifurcation of the nesting features can arise only due to nonlinearity in the dispersion which causes more than one configuration for the dispersion to match, resulting in more than one bias voltage where nesting occurs for a given $\theta$. Instead of a prolonged nesting condition, one obtains two short "kissing" lines, at which the bands of tip and sample run approximately parallel only for a short range in *k*-space (see Fig 1h). Similarly, the nonlinearity in the onset condition is impossible if the dispersion is perfectly linear.

We call these lines "nesting I" and "nesting II" as is shown in the schematic in Fig. 1h, where "nesting I" occurs when the tip and sample dispersions cross at each other's Fermi level, and "nesting II" occurs when the dispersions cross between the Dirac points and the local velocities are equal. These two angles are the same if the Fermi velocity is constant[1] but are different, leading to splitting, for a nonlinear dispersion like that shown in Fig 1e. In principle, the two nesting lines should exist also for lower bias values, however in this limit they run very close to one another and cannot be resolved. At high bias the two lines separate substantially and can be resolved, as is clearly seen in Fig. 2b. Since "nesting I" occurs at an energy $\mu$ above the Dirac point and "nesting II" occurs at an energy $\Phi/2$ below the Dirac point (see SI), the splitting of the nesting lines measures the change in the Fermi velocity over a large energy range of $\mu + \Phi/2 = V_b/2$, magnifying the effect of any nonlinearity. This unique measurement, distinct from SdH, ARPES and capacitance probes, avoids the difficulties of charge inhomogeneity that are exacerbated at the Dirac point.

**<u>Investigating the origins of the non-linear dispersion</u>**

In the following we theoretically investigate the origin of the non-linear dispersion and model its effect on the QTM spectra. The simplest possible model of the graphene band-dispersion is the single-particle model consisting of nearest neighbor hopping (NN) between graphene's $p_z$ orbitals. This model predicts a linear Dirac dispersion close to the K and K'-points. However, at higher energies the dispersion strongly deviates from linearity due to periodicity on the Brillouin zone even in the NN model. This happens near the M-points at around $\mu \sim 1.5$ eV, where also the Fermi surface become trigonally warped[27]. The band structure can however also be distorted at much lower energies e.g. by strain[28] and next-nearest neighbor hopping (NNN)[29]. In Fig. 3 we investigate *all* of these cases, where we plot the full band structures, their dispersion along the twisting path $k_\theta$, and the numerically calculated QTM spectra for the experimentally accessible range of $\theta$ and $V_b$ (see SI for details of the theoretical model). Our linear response calculations take into account the tight-binding band structure and the self-consistent electrostatics in the graphene-hBN-graphene junction. The electrostatic model requires the junction's areal capacitance $\epsilon_\perp/d = 2.4\epsilon_0/nm$ which we fit from the small-bias data (see SI).

Fig. 3a-c show the results for the simplest case, where we only include an effective NN hopping term with $t = 3.2$eV (or $v_F = 1.04 \times 10^6 m/s$) in the single particle band structure. The band structure in the K-points along the twisting direction (Fig. 3b) remains almost perfectly linear up to the experimentally accessible Fermi energy of $\mu \sim 0.35$ eV which is achieved for the maximal bias voltage of $V_b = \pm 1.5$ V (see electrostatic simulations in the SI). The corresponding QTM spectrum in Fig. 3c clearly produces the linear onset and a single nesting line, as expected from a linear Dirac dispersion, and does not show any signatures of a non-linear onset or split nesting condition. As the Fermi energy that is probed here is below the non-linearities that occur at around $\mu \sim 1.5$ eV (Fig. 3a), the QTM spectrum in our experimental range is not altered by the small trigonal warping (see SI for QTM spectra in a larger voltage range).

Fig. 3d-f show the results if we include additional -0.5% bi-axial strain in the band structure of the tip. This small value is consistent with normal graphene Raman spectra[30]. The band structure (Fig. 3d) remains almost perfectly linear close to the K-points, but develops a gap along the QTM momentum space path (see Fig. 1d) since strain shifts the tip's Dirac point off the sample's K point. This is clearly reflected in the corresponding QTM spectrum in Fig. 3f, which shows a linear onset and a single nesting line at higher bias, but exhibits a gap-like feature close to zero bias, resulting in a distinctly different spectrum than the experimental one. We have also performed the same exercise for uniaxial strain (see SI), where we find that rotation symmetry breaking leads to tripling of the nesting line and is inconsistent with the experimental data. Since the splitting in the nesting condition is highly reproducible among different devices, tips and sample locations, we conclude that a strain-driven mechanism of these features is highly unlikely.

Fig. 3g-i shows the results if we include NNN in addition to the NN hopping term in the single particle band structure using $t' = 0.3$ eV and keeping $t = 3.2$eV. Here we use the highest value of NNN hopping $t'$ that has been previously reported in experiments[31]. The band structure (Fig. 3g) becomes electron-hole asymmetric and gives rise to slight non-linear corrections close to the K-points[31–33], which are clearly visible along the twisting direction in Fig. 3h. However, these corrections are not large enough to significantly alter the QTM spectrum in Fig. 3f, which, within the accessible bias voltages, shows onset and nesting features that are very similar to the NN-only QTM spectrum. While above $V_b > \pm 1.5$ V we start to observe hints of a splitting of the nesting line (see also QTM spectra with larger voltage range in the SI), it however does not fit the experimentally observed splitting in two key ways. First, the experiment shows splitting already at $V_b \sim 1$ V which is too small a bias to be reproduced with a realistic NNN hopping. Second, we find (see SI) that the general line-shape of the splitting, where the lower branch splits off from the upper branch, is reversed as compared to the experiment.

**Renormalized Dirac cones due to electron-electron interactions**

The introduction of electron-electron interactions, which is neglected in the single-particle models above, fully resolves this disagreement. As long-range Coulomb interactions are weakly screened at the charge neutrality point due to the vanishing density of states, it produces a self-energy correction that leads to an enhancement of the Fermi velocity and a corresponding deviation from the bare linear Dirac dispersion. Many techniques, including the renormalization group[34], lattice quantum Monte Carlo[7], and Hartree-Fock[16] calculations have been employed to resolve this effect. All lead to a logarithmic correction to the dispersion relation due to Fock-like exchange interactions, which is particularly strong at the charge neutrality point and has the following analytical form:

$$E(k) = \pm \hbar v_0 k \left(1 + \frac{\alpha}{4} \ln\left(\frac{\Lambda'}{\hbar v_0 k}\right)\right) \quad \text{Eq. 1}$$

Where $v_0 = 0.823 \times 10^6$ m/s the bare velocity corresponding to a NN hopping of t = 2.5eV, $\alpha = e^2/(4\pi \epsilon_\parallel \hbar v_0)$ is the effective fine structure constant of graphene, and $\Lambda'$ is an energy cutoff of the order of the bandwidth that accounts for the breakdown of the Dirac description at high energies, $\epsilon_\parallel$ is the effective in-plane dielectric constant of graphene on hBN, $k$ is the wave vector and $\hbar$ is Planck's constant.

Within the continuum Hartree-Fock approximation, we analytically compute $\Lambda' = 4\Lambda/\sqrt{e}$ to leading order in $k$ where $\Lambda/v_0$ is the hard momentum space cutoff. The value of $\Lambda$ cannot be obtained within the continuum approximation and theoretically must be computed from detailed microscopic calculations. Accordingly, we perform self-consistent Hartree-Fock calculations incorporating screening via the random phase approximation starting from the *ab initio*-derived interacting tight binding model including long-range hopping beyond NNN terms[16]. The resulting bands are plotted in Fig. 4a (details will be given in a separate work) As we show in Fig. 4b, the numerical dispersion is quantitatively matched by Eq.1 with the value $\Lambda \sim 3.3$ eV. Remarkably, we find that fitting $\Lambda$ to the experimental data using QTM spectrum computed with Eq.1 can successfully reproduce the splitting of the nesting lines with a value of $\Lambda = 3.3$ eV. The resulting QTM spectrum is shown in Fig. 4c and matches the experimental spectrum qualitatively and quantitively. It replicates well the low and high bias features, in particular the non-linear onset line and the splitting of the nesting line. The interacting dispersion successfully reproduces the line-shape of the splitting, where the upper nesting II branch separates from the continuous nesting I branch, and become stronger with increased bias. We find that the nonlinear deviations from a constant Fermi velocity in the band structure necessitated by the split nesting condition can be accounted for entirely by interactions.

**Fitting and extraction of parameters**

To highlight the perfect match between the e-e theory and the experimental QTM spectrum, we overlay the data with analytical calculations of the generalized onset and nesting features in Fig. 4d. The black dashed lines represent the analytical solution for momentum-energy resolved tunneling with the interacting dispersion in Eq. 1 for the realistic model parameters given in Fig. 4 (see SI for expressions), whereas the red dashed lines represent the analytical solution with a linear Dirac dispersion and the effective velocity $v_F = 1.04 \times 10^6$ m/s. Deviations from the constant $v_F$ model appear in the onset feature at larger bias, and the splitting of the nesting lines matches the data to high accuracy.

Employing the high reproducibility of the splitting features, we can also statistically cluster positions of the nesting lines in the various measurements taken on the same sample A - therefore the same $\epsilon_\perp/d$ - and combine the average position, including error bars, of these in Fig. 4e. We numerically fit this cumulative nesting data with the Fock dispersion in Eq. 1[4,34], and plot it as a black fitting line. Here we use the following parameters $v_0 = 0.823 \times 10^6$ m/s, $\epsilon_\perp/d = 2.4$ /nm, and $\Lambda = 3.3$ eV. We extract graphene's fine structure constant α to be $\alpha = 0.32 \pm 0.01$ and $\epsilon_\parallel = 8.3 \pm 0.3$, which fits well within the range of previously reported values, which varied for α between 0.05 and 1.43 and for $\epsilon_\parallel$ between 2.33 and 44 in the various used dielectric environments such as vacuum, hBN and SrTiO$_3$[4–6,11,12,18,19,20]. The extracted parameters compare extremely well with previous reports on low temperature measurements of hBN-encapsulated devices, where an $\alpha = 0.275$, $\epsilon_\parallel = 8$ and $v_0 = 0.85 \times 10^6$ m/s were reported[12].

**Conclusion**

Our study uncovers the exceptional capabilities of the Quantum Tunneling Microscope (QTM), which through its interferometric tunneling mechanism enhances sensitivity to the faintest modifications in the electronic bands of two-dimensional materials, capturing subtle effects that are often invisible to conventional probes. This enables direct observation of finely detailed, interaction-driven band structures, even at room temperature, offering remarkable reproducibility.

Our results highlight pronounced electron-electron interaction effects, even in symmetric, non-ordered phases of graphene. They further establish the QTM as a uniquely powerful tool for mapping spectral functions of strongly correlated quantum states, like superconductivity[31], correlated insulators[32] and fractional Chern insulators[33], in a wide variety of systems such as twisted and untwisted graphene and transition metal dichalcogenides (TMDs) compounds.

**Acknowledgements**

The authors thank Shahal Ilani and Alon Inbar for a fruitful discussion regarding the building and operation of the QTM and Leonid Glazman and Felix von Oppen for discussions concerning theoretical modelling. M.L acknowledges funding from Munich Center for Quantum Science and Technology (MCQST) Instrumentation and Seed funds. D.K.E. acknowledges funding from the European Research Council (ERC) under the European Union's Horizon 2020 research and innovation program (grant agreement No. 852927), the German Research Foundation (DFG) under the priority program SPP2244 (project No. 535146365), the EU EIC Pathfinder Grant "FLATS" (grant agreement No. 101099139) and the Tschira foundation under the project SuperC. JHA is supported by a Hertz Fellowship. B.A.B. was supported by the Gordon and Betty Moore Foundation through Grant No. GBMF8685 towards the Princeton theory program, the Gordon and Betty Moore Foundation's EPiQS Initiative (Grant No. GBMF11070), the Office of Naval Research (ONR Grant No. N00014-20-1-2303), the Global Collaborative Network Grant at Princeton University, the Simons Investigator Grant No. 404513, the BSF Israel US foundation No. 2018226, the NSF-MERSEC (Grant No. MERSEC DMR 2011750), the Simons Collaboration on New Frontiers in Superconductivity (SFI-MPS-NFS00006741-01), and the Schmidt Foundation at the Princeton University. This project was partially supported by the European Research Council (ERC) under the European Union's Horizon 2020 Research and Innovation Program (Grant Agreement No. 101020833).

# References

1. Inbar, A. *et al.* The quantum twisting microscope. *Nature* **614**, 682–687 (2023).

2. Birkbeck, J. *et al.* Quantum twisting microscopy of phonons in twisted bilayer graphene. *Nature* 641, 345–351 (2025).

3. Klein, D. R. *et al.* Imaging the Sub-Moir\'e Potential Landscape using an Atomic Single Electron Transistor. *arXiv:2410.22277* (2024).

4. Elias, D. C. *et al.* Dirac cones reshaped by interaction effects in suspended graphene. *Nat. Phys.* **7**, 701–704 (2011).

5. Hwang, C. *et al.* Fermi velocity engineering in graphene by substrate modification. *Sci. Rep.* **2**, 590 (2012).

6. Ryu, H. *et al.* Temperature-Dependent Electron–Electron Interaction in Graphene on $SrTiO_3$. *Nano Lett.* **17**, 5914–5918 (2017).

7. Tang, H.-K. *et al.* The role of electron-electron interactions in two-dimensional Dirac fermions. *Science* **361**, 570–574 (2018).

8. Kotov, V. N., Uchoa, B., Pereira, V. M., Guinea, F. & Castro Neto, A. H. Electron-Electron Interactions in Graphene: Current Status and Perspectives. *Rev. Mod. Phys.* **84**, 1067–1125 (2012).

9. Chen, X. *et al.* Electron-electron interactions in monolayer graphene quantum capacitors. *Nano Res.* **6**, 619–626 (2013).

10. Siegel, D. A., Regan, W., Fedorov, A. V., Zettl, A. & Lanzara, A. Charge-Carrier Screening in Single-Layer Graphene. *Phys. Rev. Lett.* **110**, 146802 (2013).

11. Siegel, D. A. *et al.* Many-body interactions in quasi-freestanding graphene. *Proc. Natl. Acad. Sci.* **108**, 11365–11369 (2011).

12. Yu, G. L. *et al.* Interaction phenomena in graphene seen through quantum capacitance. *Proc. Natl. Acad. Sci.* **110**, 3282–3286 (2013).

13. Lucas, A. & Fong, K. C. Hydrodynamics of electrons in graphene. *J. Phys. Condens. Matter* **30**, 053001 (2018).

14. Sonntag, J. *et al.* Impact of Many-Body Effects on Landau Levels in Graphene. *Phys. Rev. Lett.* **120**, 187701 (2018).

15. Hirata, M., Kobayashi, A., Berthier, C. & Kanoda, K. Interacting chiral electrons at the 2D Dirac points: a review. *Rep. Prog. Phys.* **84**, 036502 (2021).

16. Stauber, T. *et al.* Interacting Electrons in Graphene: Fermi Velocity Renormalization and Optical Response. *Phys. Rev. Lett.* **118**, 266801 (2017).

17. Hofmann, J., Barnes, E. & Das Sarma, S. Why Does Graphene Behave as a Weakly Interacting System? *Phys. Rev. Lett.* **113**, 105502 (2014).

18. Jobst, J., Waldmann, D., Gornyi, I. V., Mirlin, A. D. & Weber, H. B. Electron-Electron Interaction in the Magnetoresistance of Graphene. *Phys. Rev. Lett.* **108**, 106601 (2012).

19. Walter, A. L. *et al.* Effective screening and the plasmaron bands in graphene. *Phys. Rev. B* **84**, 085410 (2011).

20. Chae, J. *et al.* Renormalization of the Graphene Dispersion Velocity Determined from Scanning Tunneling Spectroscopy. *Phys. Rev. Lett.* **109**, 116802 (2012).

21. Xiao, J. *et al.* Theory of phonon spectroscopy with the quantum twisting microscope *Phys. Rev. B* **110**, 205407 (2024).

22. Wei, N. *et al.* Dirac-point spectroscopy of flat-band systems with the quantum twisting microscope. *Phys. Rev. B* **111,** 085128 (2025).

23. Wei, N. *et al.* Theory of plasmon spectroscopy with the quantum twisting microscope. *arXiv:2506.05485* (2025).

24. Peri, V. *et al.* Probing quantum spin liquids with a quantum twisting microscope. *Phys. Rev. B* **109**, 035127 (2024).

25. Luican, A., Li, G. & Andrei, E. Y. Quantized Landau level spectrum and its density dependence in graphene. *Phys. Rev. B* **83**, 041405 (2011).

26. Zhang, Y., Tan, Y.-W., Stormer, H. L. & Kim, P. Experimental observation of the quantum Hall effect and Berry's phase in graphene. *Nature* **438**, 201–204 (2005).

27. McCann, E. & Koshino, M. The electronic properties of bilayer graphene. *Rep. Prog. Phys.* **76**, 056503 (2013).

28. Guinea, F., Katsnelson, M. I. & Geim, A. K. Energy gaps and a zero-field quantum Hall effect in graphene by strain engineering. *Nat. Phys.* **6**, 30–33 (2010).

29. Kadirko, V., Ziegler, K. & Kogan, E. Next-Nearest-Neighbor Tight-Binding Model of Plasmons in Graphene. *Graphene* **2**, 97–101 (2013).

30. Ferralis, N. Probing mechanical properties of graphene with Raman spectroscopy. *J. Mater. Sci.* **45**, 5135–5149 (2010).

31. Kretinin, A. *et al.* Quantum capacitance measurements of electron-hole asymmetry and next-nearest-neighbor hopping in graphene. *Phys. Rev. B* **88**, 165427 (2013).

32. Suprunenko, Y. F., Gorbar, E. V., Loktev, V. M. & Sharapov, S. G. Effect of next-nearest-neighbor hopping on the electronic properties of graphene. *Low Temp. Phys.* **34**, 812–817 (2008).

33. Ahmed, M. The next nearest neighbor effect on the 2D materials properties. Preprint at https://doi.org/10.48550/arXiv.1111.0104 (2011).

34. González, J., Guinea, F. & Vozmediano, M. A. H. Non-Fermi liquid behavior of electrons in the half-filled honeycomb lattice (A renormalization group approach). *Nucl. Phys. B* **424**, 595–618 (1994).

35. Cao, Y. *et. al*. Unconventional superconductivity in magic-angle graphene superlattices. *Nature* 556, 43–50 (2018).

36. Cao, Y. *et. al*. Correlated insulator behaviour at half-filling in magic-angle graphene

superlattices. *Nature* 556 (7699), 80-84 (2018).

37. Park, H. *et. al*. Observation of fractionally quantized anomalous Hall effect. *Nature* 622, 74–79 (2023).

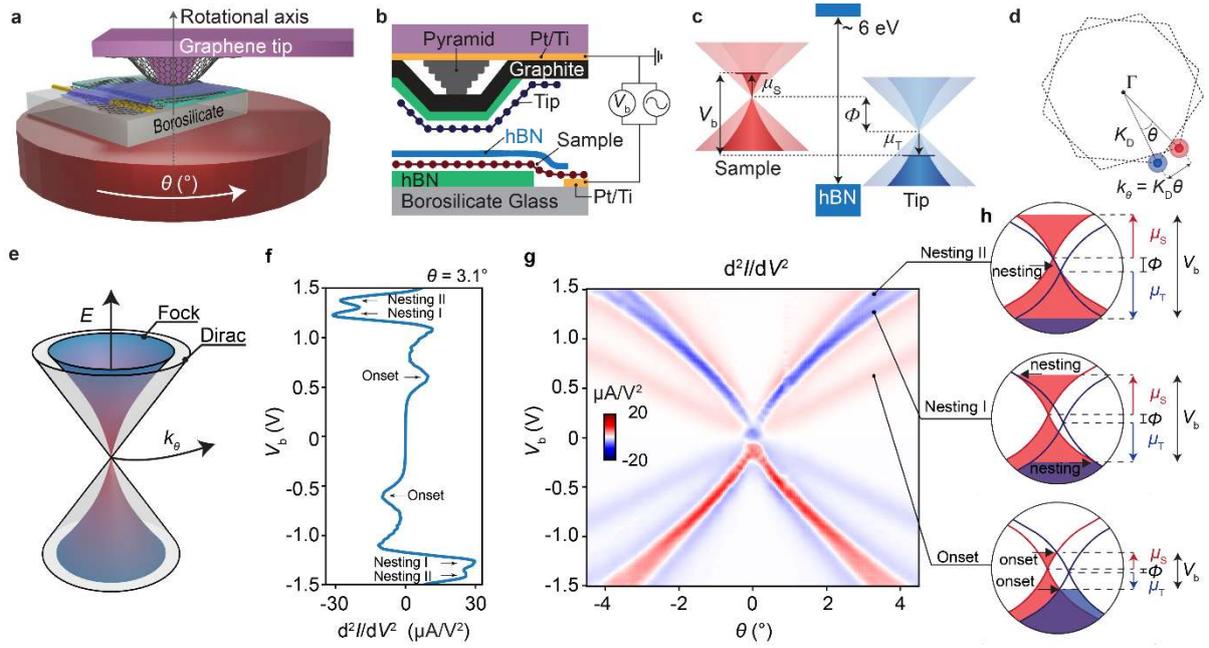

**Figure 1. Working principle of the QTM and measurements of the twist-angle dependent graphene-hBN-graphene tunneling spectra. a-b,** Illustration of the QTM setup which uses a graphene-coated pyramid on a cantilever, and a hBN/graphene-coated substrate that is mounted on a rotating platform. D.C. and A.C. signals are biased on the sample and the resulting tunneling current is collected by the tip. **c,** Tunneling spectroscopy across the graphene/hBN/graphene interface. Biasing the sample with $V_b$ shifts the potential $\phi$ as well as the chemical potentials of the graphene samples $\mu_S$ and tip $\mu_T$ while preserving over-all charge neutrality. **d,** Rotated Brillouin zones and Fermi surfaces of the sample and tip. The relative distances in k-space between the Dirac points can be estimated by $k = K_D \theta$. **e,** Illustration showing the difference in the band structure of graphene between the interaction driven nonlinear dispersion (Fock) and the linear dispersion without interaction (Dirac). **f-g,** Experimental and theoretical elastic tunneling spectroscopy as a function of relative twist-angle $\theta$ and bias voltage $V_b$, showing the signatures of onset and nesting conditions. **f,** Line trace of the derivative of the differential conductance ($d^2I/dV^2$) and bias voltage $V_b$ at a twist angle of $\theta = 3.1°$ which shows peaks and troughs associated to the onset and nesting conditions. Nesting features at high $V_b$ are split into Nesting I and Nesting II. **g,** 2D $d^2I/dV^2$ tunneling map with respect to the twist angle $\theta$ and $V_b$ with three distinct features "Onset", "Nesting I" and "Nesting II", highlighted with illustrations in **h**. Two nesting lines can occur if the bands have nonlinearity in their dispersion.

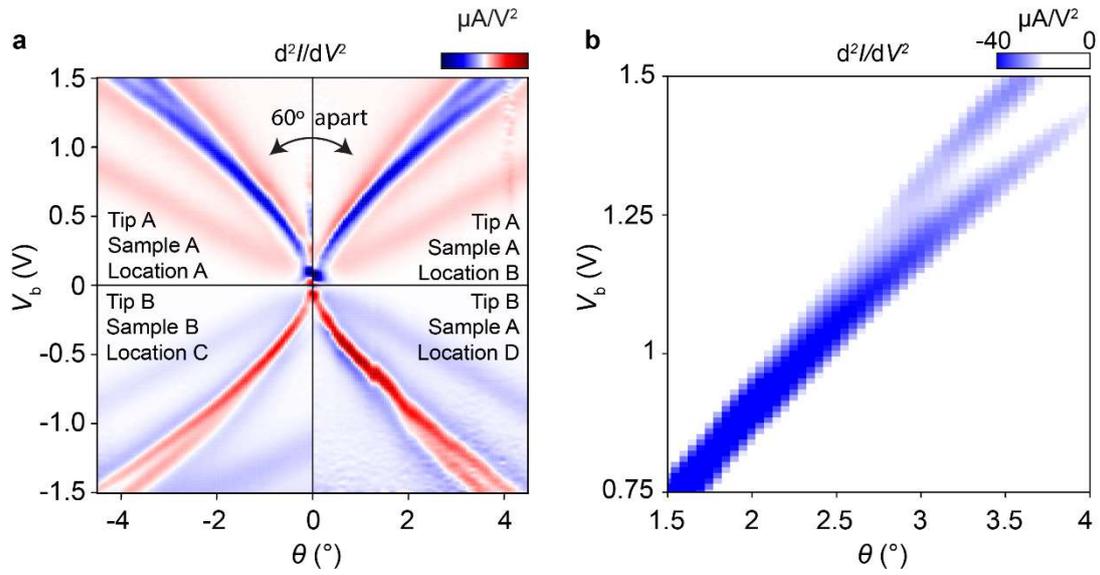

**Figure 2. Angle resolved tunneling spectroscopy of graphene-hBN-graphene showing reproducibility of the splitting feature. a**, $d^2I/dV^2$ map from four independent data sets arising from permutations of different tips, samples and locations on the sample. Top two quadrants are data taken from two different real space locations and 60° apart in $k$-space. Bottom two quadrants are data taken on two different samples with the same tip. **b**, $d^2I/dV^2$ map zoomed-in near the splitting event showing the Nesting I branch stemming off from the Nesting I branch.

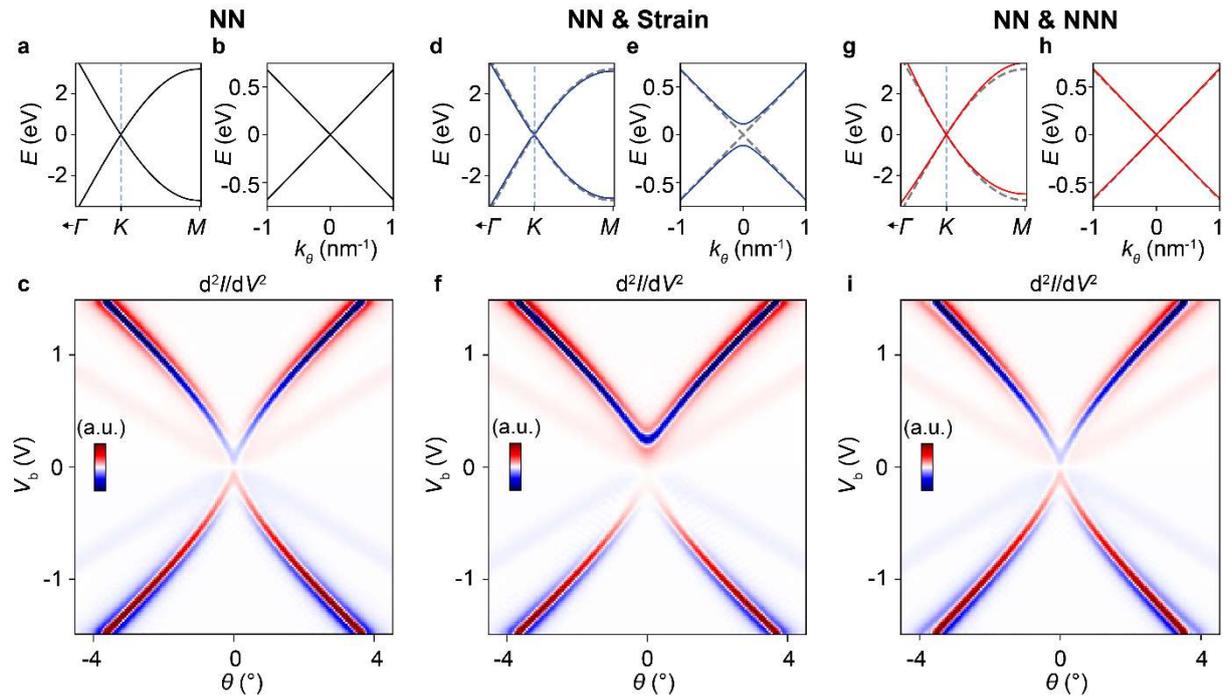

**Figure 3. Exploring the origins of the nesting split. a,** The band structure of graphene, **b,** dispersion of graphene along the twisting path - $k_\theta$ and **c,** a simulated conductance $d^2I/dV^2$ map for monolayer graphene with only the nearest-neighbor (NN) hopping considered. **d-f,** is the same, with NN and -0.5% biaxial strain and **g-i,** with NN and next-nearest neighbor (NNN) included in the simulation. The dashed grey lines in **d, e, g, h,** show the dispersion for the NN case. We use t = 3.2eV for the NN hopping and t' = 0.3eV for the NNN hopping.

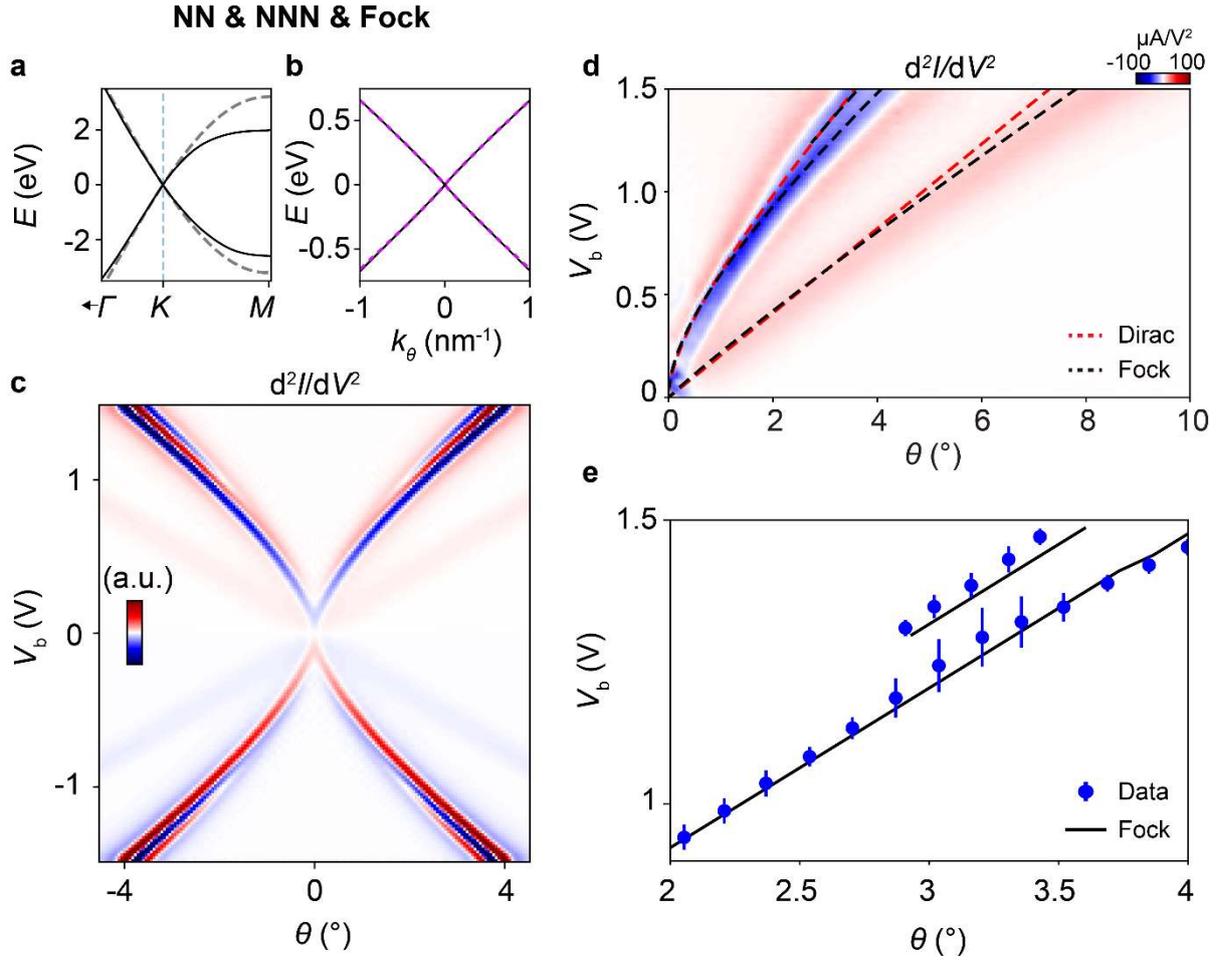

**Figure 4. Introduction of the e-e interaction in the model reproducing the splitting features. a,** The numerical Hartree-Fock band structure (black) computed using a ab initio-derived tight-binding model. For reference, the NN+NNN model of Fig. 3a is shown in gray, **b,** Zoom-in of 4a along the twisting path - $k_\theta$ compared to Eq. [1] with $\Lambda = 3.3$eV, showing excellent agreement, **c,** resulting simulation for the $d^2I/dV^2$ of the Fock-renormalized dispersion in Eq. [1]. We use $\Lambda = 3.3 eV, \alpha = .32, \epsilon_\perp/d = 2.4\epsilon_0\text{nm}^{-1}$, **d,** A dataset taken over a wide angular range, with analytical nesting and onset lines plotted in dashed red (assuming linear Dirac dispersion with $v_F = 1.04 \times 10^6 m/s$) and black (assuming the model in **a-c**). The nonlinearity of the onset as well as the splitting features are better captured by the interacting model. **e,** Statistical clustering of 7 of our datasets taken from a single sample with different tips (blue points with error bars). The peaks from the numerical model shown in **c**, are extracted and plotted as solid black lines. Best-fit parameters are extracted by a least-squares optimization of $\epsilon_\perp/d, \alpha$, and $\Lambda$ with the bare velocity $v_0 = 0.82 \times 10^6$m/s held fixed.

# Supplementary information:

# Experimental methods

E1: QTM setup and measurement scheme

QTM setup: we have adapted a commercial AFM (Veeco Dimension 3100) to facilitate our QTM experiments, following a similar protocol as in the original QTM work [1]. As is shown in Fig. 1, the original sample stage is replaced with a stack of X-Y-θ positioners that are inclined on an aluminum wedge of ~ 11°. The complete motorized stage consists of several pieces (from bottom to top): (i) 11-degree (°) wedge plate: which is fixed to the X-Y motors of the AFM for the alignment of the sample with respect to the pyramid on the cantilever. In the AFM that we are using, the cantilever is mounted at an angle ~10 - 12°. To compensate for this tilt angle, we have added a ~ 11° wedge shaped metal plate under all the piezo stages. This reduces the relative tilt between the tip and the sample and allows to make a nearly parallel contact between the QTM tip and the sample. (ii) Rotational stage: a Thorlabs direct-drive rotational stage (DDR100/M controlled with BBD301) is mounted on top of the 11° wedge plate to make sure that the sample is rotating parallel to the tip. The mass of the DDR100/m, the ultra-low wobble, the high-resolution encoder (2 µrad) and its smooth continuous movements are ideal for QTM measurements. (iii) X-Y stage: is mounted on top of the rotational stage to correct the sample position with respect to the rotational axis. Here we use Thorlabs (M30XY/M) which provide < 1 µm translational travel accuracy. These two stages along with the rotator are used to position the device at the center of rotation. (iv) Sample holder: we use a custom-built PCB to wire-bond our devices. A Teflon plate in between the stages and the sample holder acts as an isolator for the X-Y-θ stages in case of potential electrical crosstalk with the sample. The sample holder sits on an elevated platform, which allows us to make sure that the chip is residing on a higher plane and the tip can touch the tunneling barrier while rotating the sample without hitting any other component of the stage.

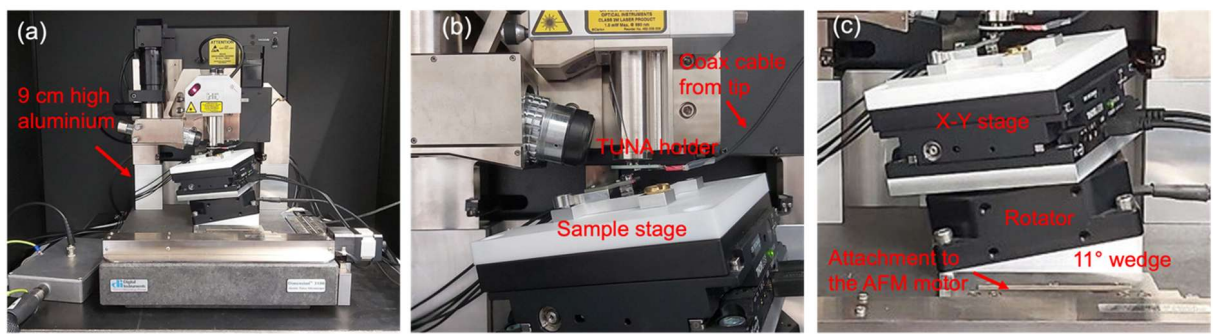

Fig.1: Image of the QTM setup. (a) Full setup showing the aluminum plate used to raise the scan head. (b) Interface of the sample and the tip along with the coax cable for readout. (c) Full stack of the X-Y-θ stages and the 11° wedge.

QTM measurement scheme: to engage the tip with the sample we use a commercial scan head. All our stages for the device side have a total height of ~ 9 cm. Hence, to fit these stages, we use a ~ 9 cm high aluminum block to raise the AFM scan head (Fig. 1a). A coaxial cable is soldered onto the Bruker DTRCH-AM (TUNA) tip holder to electrically connect the tip. This cable is connected to the final breakout box. Our TUNA holder has one electrical connection since in all our measurements we varied the bias voltage without applying any gate voltage to the tip.

We employ integrated optics and standard AFM operations to bring the two van der Waals surfaces (located on the AFM tip and the sample) into contact of the tunneling barrier. Standard AFM operations facilitate precise control, enabling us to achieve and maintain electrical contact between the vdW interfaces. Additionally, we have used active force feedback, a technique that continuously adjusts the force applied by the tip to ensure consistent contact between the tip and the sample during rotational measurements. Furthermore, we have used a continuous contact mode. This mode involves bringing the tip and sample into contact and then rotating the sample while keeping the two surfaces in contact throughout the entire measurement process. This approach allows us to perform accurate rotational measurements without losing contact.

The AC bias voltage is applied with a SRS860 lock-in amplifier, and the DC bias voltage is applied from the Keithley 2450 source meter. Both AC and DC voltages are mixed with an SRS SIM980 summing amplifier are applied to the device on the rotation stage. On the other side, the current is collected from the tip and measured with the lock-in (AC voltage) and multimeter (Keithley DMM 7510, DC voltage) after being amplified with a Femto DLPCA-200 current amplifier.

E2: Fabrication of graphene coated QTM tips

Fabrication of QTM cantilevers: tip fabrication is a crucial step in our QTM measurements as it ensures electrical contact between the van der Waals materials on the tip and sample sides. The quality of this interface directly influences the quality of the data. As a starting point, we use a commercial tipless AFM cantilever (TL-NCL) with a spring constant of $k \sim 48$ N/m, length $l \sim 225$ μm, width $w \sim 38$ μm and a thickness of $t \sim 7$ μm. We deposit a thin film of 10 nm Ti and 40 nm Pt on the body of the cantilever, that provides a good electrical contact.

We further deposit a pyramid on the cantilevers, which acts as an elevated platform for further graphene coating. We use two different procedures to deposit the pyramids, both of which work well and produce similar results in the QTM measurements. In both procedures, we construct the pyramids with height $1 - 2$ μm, a base area of 2μm x 2μm and with a flat apex of ~ 200nm x 200nm. To facilitate the formation of planar 2D plateaus which are defined by the folds in the 2D materials as they conform to the topography of the tip, we have optimized the deposition parameters to create as flat as possible apexes of the pyramids.

The first procedure uses deposition of tungsten on the cantilever with focused ion beam induced deposition (FIBID). Typically, our pyramids consist of several consecutively deposited layers with different sizes, starting from the base to apex. Fig. 2 shows the SEM images of these pyramids. We always check the pyramids with SEM before transferring the 2D materials to make sure that the apex is flat.

In addition to FIBID, we have developed a deposition technique that produces diamond-like carbon (DLC) pyramids through e-beam induced deposition (EBD) using oil as a precursor [2]. The DLC has been shown to display great mechanical strength, making it a viable alternative to FIBID deposited pyramids. Here, the high energy electron beam knocks off the long chain carbon polymer of the organic oil, so depositing carbon atoms on the cantilever, which assemble into a diamond-like crystalline structure. We have rigorously varied the key parameters in the SEM such as, acceleration voltage, aperture diameter, magnification, dwell time per pixel and total exposure time to understand their roles in the deposition process. Each of these parameters was thoroughly explored to determine the optimal conditions for depositing a multilayered pyramid structure with a flat apex. Additionally, we used conductive copper tape as shown in Fig. 3 to secure the cantilevers to the sample holder, preventing charging during deposition that could potentially misalign consecutive layers. This process creates a thin film of amorphous carbon on top of the body of the cantilever close to the pyramid. Hence, a proper cleaning of the cantilever is necessary before depositing 2D materials on top. After the DLC deposition, the cantilevers are cleaned with acetone, IPA followed by ozone (Ossila UV ozone cleaner) for 10 mins to remove the unwanted carbon deposition. After each deposition, we perform Raman spectroscopy to confirm the material's composition as carbon [3].

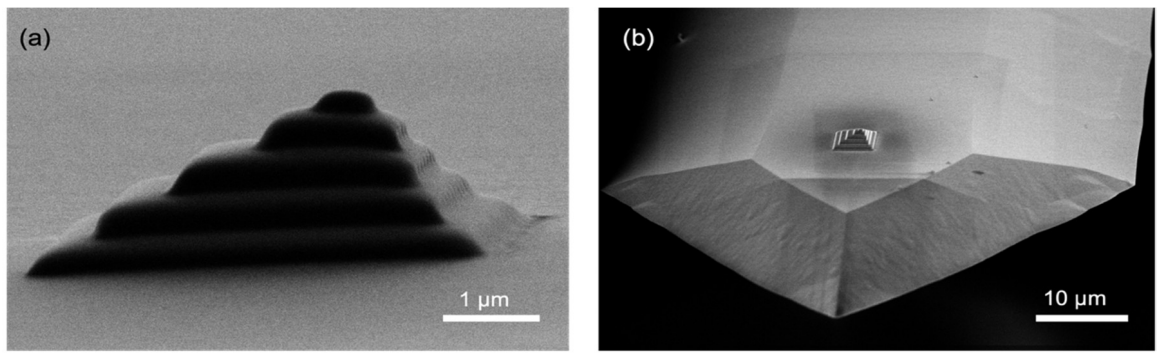

Fig. 2: SEM image (a) SEM image of the pyramid. Apex is ~ 200nm x 200nm and height is ~ 2 μm. (b) The pyramid is deposited at the center of the tipless cantilever.

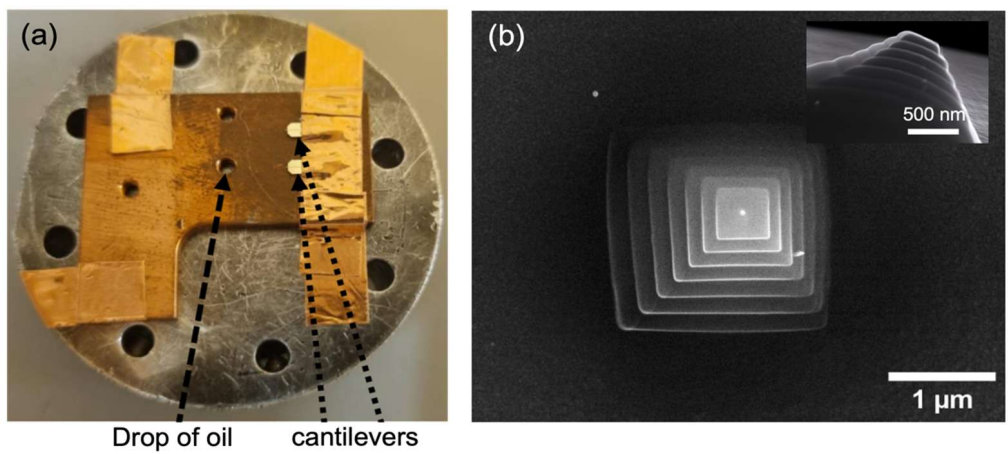

Fig. 3: (a) Sample holder with the cantilevers and oil droplets as the precursor for the EBD deposition. (b) SEM image of a DLC pyramid grown with the EBD method.

Transfer of 2D flakes on the cantilever: after the pyramid deposition on the tipless cantilever, 2D flakes are transferred onto the pyramid. We use graphene flakes to contact the samples in all our measurements. To improve the mechanical stability of the graphene flake, and to create a soft cushion, we use ~ 10 nm graphite flake and ~ 10 - 20 nm hBN flake underneath the graphene flake. We used a dry transfer technique to transfer these 2D materials on the pyramid.

Graphene exfoliation: To increase the probability of having graphene flakes on the polymer coated chips, we need to apply more pressure on the crystal containing tape than the typical graphite/hBN exfoliation. Hence, we use another adhesive sacrificial layer of polymer underneath the PPC for graphene exfoliation. First, a layer of polyvinyl pyrrolidone (PVP) is spin coated on bare Si at 4000 rpm for 40 sec using a solution of 1% PVP dissolved in IPA, followed by 1 min baking at 100°C. Then a layer of PPC (6% in anisole) is spin coated on top of the PVP at 1800 rpm for 40 sec resulting in a purple hue similar to that of the 285 nm $SiO_2$/Si typically used to enhance contrast in 2D materials. After the spin coating, we bake it at 100°C for 1 min. Using Scotch tape, graphite is exfoliated several times and finally placed in contact with the PPC surface. Then, like the method by Huang *et al.* [5], the Si along with the tape are placed on a hot plate at 100°C for 1 minute. Afterwards, the Scotch tape is slowly peeled away. WITec Raman alpha 300 is used to confirm that the flake of interest is monolayer, before proceeding with the transfer.

Graphite and hBN exfoliation: Graphite and hBN transfer on the pyramid are slightly different than graphene transfer. We exfoliate graphite and hBN on polypropylene carbonate (PPC) coated silicon. A solution of 15% PPC in Anisole is spin coated on a piece of ~ 2 cm x 2 cm Si/$SiO_2$ chip with a rotation speed of 4000 rpm for 40 seconds. The chip is then baked at 100°C for 2 mins. After the spin coating and baking, the chip should have the violet tint. This is important for the visibility of the graphite and hBN flakes that will be exfoliated in the next step. For the exfoliation of graphite, we use blue dicing tape from Ultron. We make the graphite tape beforehand and exfoliate graphite while the chip is still hot. hBN exfoliation is also done following the same process.

Preparation for the 2D flake transfer: After identifying the desired graphite, hBN and graphene flakes, they are individually transferred onto the pyramid. For the graphite and hBN flakes, a piece of a blue dicing tape with a hole punched out is placed on the PPC surface with the flake of interest in the center of the hole. Using a sharp blade, the PPC film is cut around the edge of the blue tape. The PPC film, along with the blue tape is then carefully peeled off from the silicon substrate and attached to a metal plate with a hole punched out using double sided tape as shown in Fig. 4. These steps are repeated for the transfer of graphene, but instead of peeling the blue tape along with the PPC away from PVP, water droplets are dispensed using an Eppendorf pipette to undercut the water soluble PVP layer beneath the PPC.

Glass slide assembly for the transfer: The metal piece with the blue tape, PPC membrane and the flake of interest in the center of the hole is attached to a glass slide and mounted onto a set of X-Y-Z stages used for the 2D assembly. Meanwhile, a piece of silicon with a few layers of polydimethyl-siloxane (PDMS) is used to mount the cantilever onto the transfer stage. Using the micromanipulators, the 2D flake is positioned on top of the pyramid.

By heating up the sample stage to 120°C, well-beyond the glass transition temperature of PPC ($T_g \sim 45$°C), the membrane is melted onto the cantilever, releasing the 2D material onto the pyramid. The thick layer of PPC residue is cleaned in acetone and IPA and the steps are repeated for subsequent layers of 2D flakes.

Transfer of the 2D flake: Once the 2D flakes are prepared on the metal holder, we use the typical dry transfer technique to transfer the graphite, hBN and graphene flakes on the pyramid consecutively. After transferring each flake, the tip is cleaned with acetone and IPA to remove the PPC on top. Fig. 5 shows the optical images of multiple graphite and graphene covered tips which have been used for the measurements.

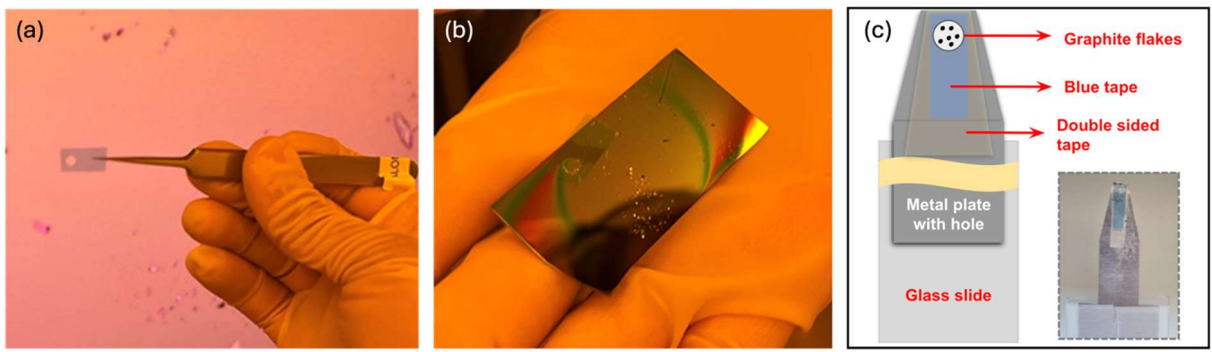

Fig. 4: Glass slide assembly for transfer. (a) small piece of the blue tape with a hole in the middle. (b) The hole in the blue tape is aligned with the flake of interest on the chip. PPC is peeled off from the chip. (c) Schematics of the glass slide assembly for the transfer on the pyramid. Optical image of such an arrangement (inset).

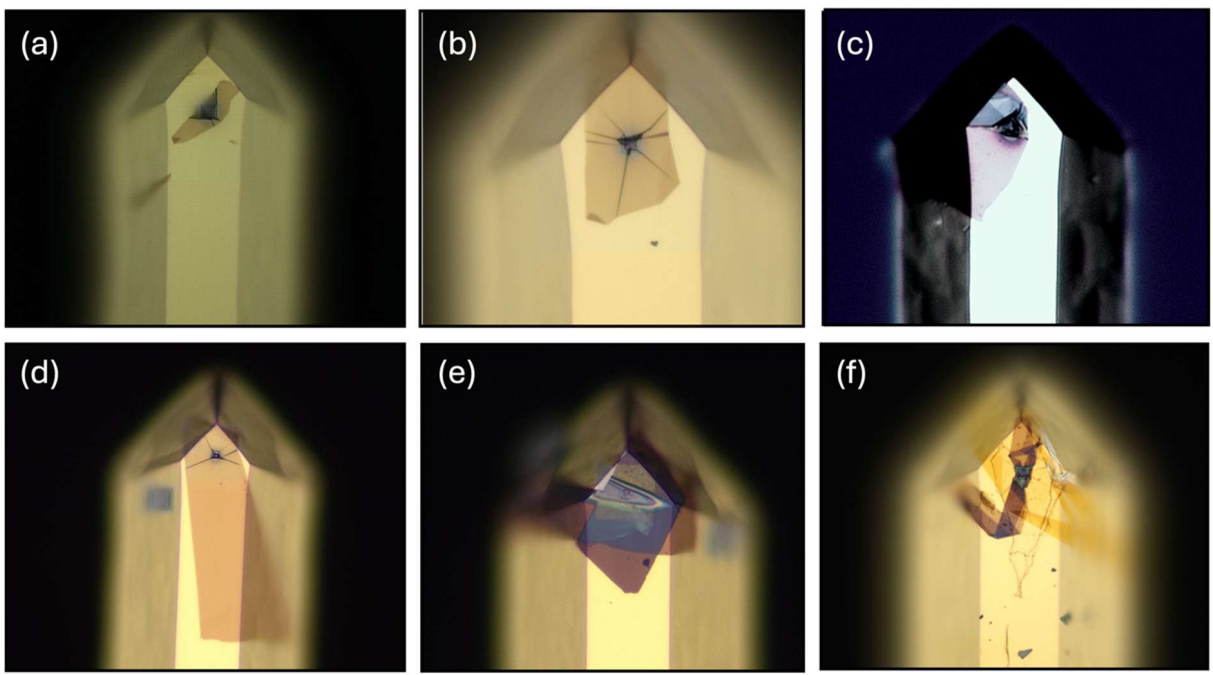

Fig. 5: Optical images of the clean tip with 2D flakes. (a), (b), (c), (d) Graphite tips. (e), (f) Graphene tips.

AFM characterization: we place the tip on top of a Si/SiO2 chip with a piece of PDMS on top for the ease of performing AFM. We image the tip with tapping mode, contact mode and peak force tapping mode AFM in order to measure the area of the plateau of the tip. Fig. 6a shows the topography of a graphene tip with graphite/hBN/MLG on top. From the height profile we can clearly see the tent on the pyramid. Fig. 6b shows the zoomed-in AFM image of the apex of the pyramid. The flake spontaneously forms a tent with a flat plateau of 350nm x 350nm in size.

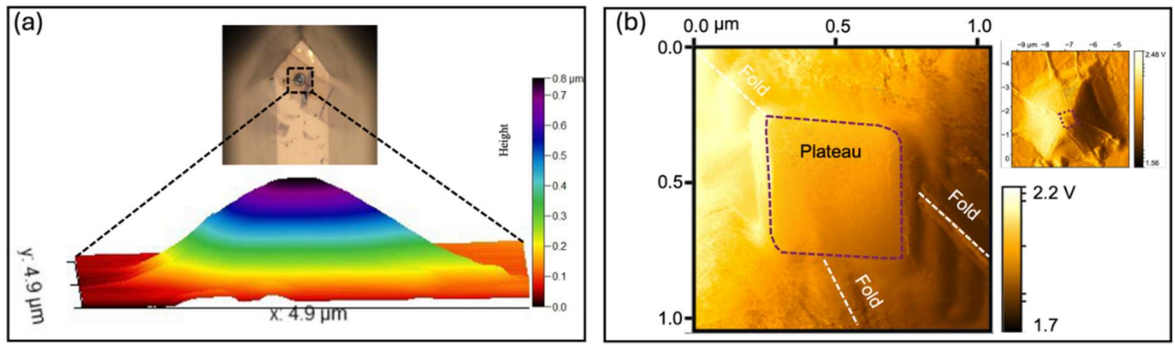

Fig. 6: AFM of the pyramid with 2D flakes. (a) Topography of the pyramid with the 2D flakes. (b) Top view of the pyramid showing the flat apex.

Raman characterization: typically, our graphene tips are consisting of graphite, hBN and graphene flakes from bottom to top. To make sure that the tip area is covered with graphene, we perform Raman measurements on the graphite flake slightly away from the pyramid (magenta circle) and on top of the pyramid (green star). Fig. 7 shows two such spectra of a graphene and a graphite region of the tip.

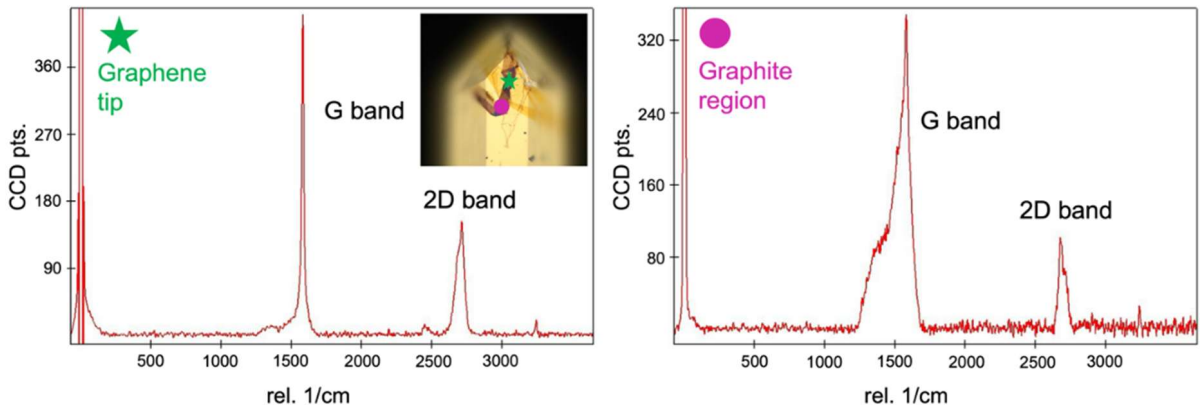

Fig. 7: Raman spectrum of the graphene and a graphite region on a graphite/hBN/graphene tip.

SEM characterization: is performed to image the entire graphene coated pyramid structure, as seen in Fig. 8. However, using SEM on the tip carries the risk of depositing amorphous hydrocarbons onto the graphene flake, which can create a thin insulating layer. To mitigate this, we only performed SEM on a few tips as sacrificial samples to understand the tip shape.

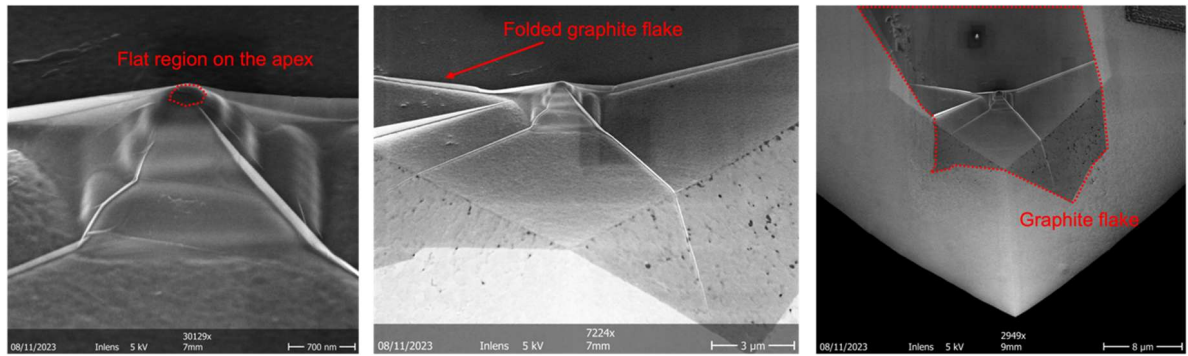

Fig. 8: SEM image of the graphite flake on the pyramid

E3: Device fabrication on the rotating substrate

Substrate fabrication: in QTM experiments, it is crucial for the device to be positioned on a raised platform to prevent the cantilever body from hitting the substrate, allowing the tip to rotate freely while maintaining contact with the device. To facilitate this, we place our devices at the extreme corner of the substrate, enabling the tip to rotate freely by more than ~ 180°, which is sufficient for our measurements. To mitigate reflection of the AFM laser beam from the substrate, we use a transparent borosilicate glass substrate, which also enhances the visibility of the 2D flakes and pre-patterned electrodes.

We begin by fabricating corner electrodes on the glass substrate, as shown in Fig. 9. 10mm x 10mm borosilicate glass are spin coated with LOR 3B at 4000 rpm, and baked at 190 C for 5 min. Then, S1813 diluted with AZ EBR (1:1) is spin coated at 4000 rpm and subsequently baked at 110 C for 1 min. The device design, arranged in 2 x 2 radially from the center as shown in Fig. 9, is exposed using µMLA, developed with AZ 726 MIF for 40 s and rinsed in DI water. Prior to metallization, the design is descummed with $O_2$ plasma (Sentech ICP-RIE SI500) to ensure proper adhesion and flat topography of the electrodes. Using e-beam evaporation (Bestec/Balzers), the electrodes are metallized with 5nm Cr and 15nm Pt. For electrical contacts to the graphite or graphene devices, we used a thin Pt layer instead of Au because our devices undergo subsequent annealing process which is incompatible with Au electrodes. Annealing is performed to ensure a clean surface after stacking different 2D layers, which ensures an effective removing of any polymer residues. The detailed process will be discussed later in the device fabrication section. After the evaporation of contacts, the metal layer is lifted off in acetone and rinsed in IPA. Prior to cleaving the substrate into individual 5mm x 5mm dies, the top side is coated with a layer of S1813 as protection. Then, using a precision glass scriber (Ultile), the back side of the substrate is scored into four quadrants and cleaved using a wafer cleaving plier. Prior to the assembly of the 2D device on the substrate, the resist is stripped using acetone and IPA and the surface is further cleaned using $O_2$ plasma using the RIE.

2D sample fabrication with ultra-clean surfaces: sample fabrication for the QTM measurements requires a lot of attention to the surface cleanliness. The devices are capped with a thin layer of hBN (1-4 layers) as the tunnel barrier to perform momentum-resolved tunneling measurements. We use standard van der Waals assembly techniques based on polycarbonate (PC)/PDMS. The device stacks are assembled and released onto the pre-patterned substrates [5]. The bulk of the PC is removed with chloroform, and the last remaining residues are cleaned by annealing the chip at 450°C in air for ~ 3 mins. For thin hBN exfoliation, we use 90 nm thick $SiO_2$/Si chips, as this thickness provides optimal visibility for hBN flakes [6]. We follow the protocol by [7] to enhance the yield of thin hBN flakes. We begin by cleaning the chips with oxygen plasma to enhance the adhesion of 2D materials to $SiO_2$. After cleaning, we attach Scotch tape with hBN crystals to the chips, press it down, and heat the chips at 100°C for 2 minutes. Following the heating, we slowly peel off the tape. Once exfoliation is complete, we carefully search for 1-4 layer hBN flakes. Despite the improved visibility provided by 90 nm $SiO_2$, identifying these thin flakes remains challenging. Ideally, we prefer to use the thin flakes attached to a slightly thicker flake to have a better visibility during the stacking process.

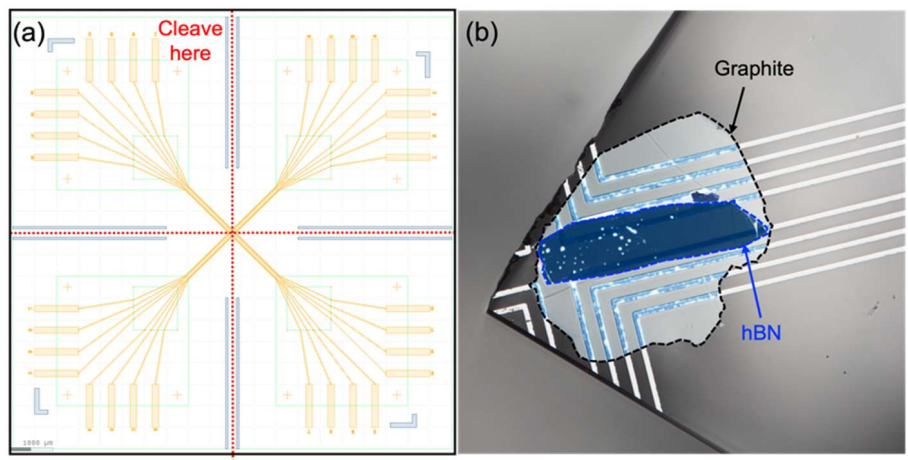

Fig. 9: 2D vdW devices on the glass substrate. (a) Design of the electrodes. (b) Optical images of graphite and hBN on the substrate.

Peak force measurements to determine hBN thickness: we measure the thickness of hBN in each device to verify the tunnel barrier thickness before performing measurements. The thickness of the tunnel barrier, *d*, has an important effect on the electrostatics of the device within theoretical calculations as described in App. T2. We have found that Peak Force Tapping mode provides significantly more reliable measurements for 2–3 layers of hBN than tapping mode. In this mode, the cantilever oscillates at a frequency much lower than its resonance frequency, driven by a sinusoidal signal to prevent unintended oscillations. As it approaches the sample, van der Waals forces attract the cantilever until contact, where Pauli repulsion causes it to deflect upwards. During retraction, the tip briefly adheres before separating, leading to a ringing motion [8]. We used a peak force of ~15 nN to extract the thickness of the hBN tunnel barrier. Fig. 10 shows the optical, AFM and Raman images of the so obtained devices. Here device A has 0.633nm thick hBN, which corresponds to bilayer hBN, whereas device B has 1.39nm thick hBN, which corresponds to tetralayer hBN.

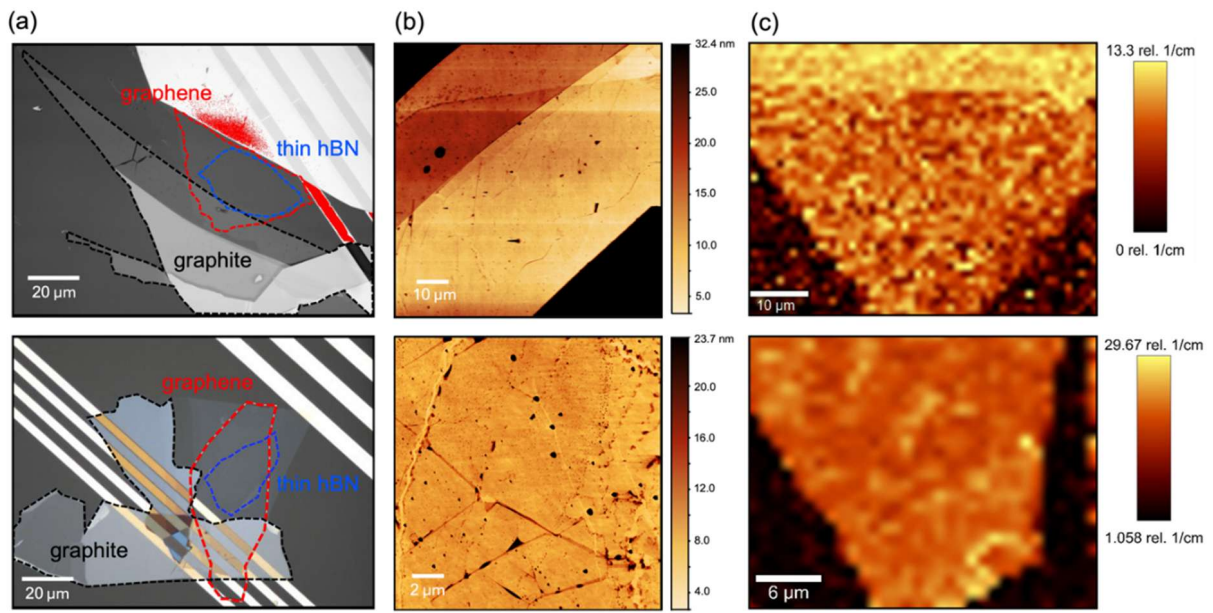

Fig 10: Shows the two monolayer graphene devices in this study. Device A is shown in the top and device B in the bottom row. (a) Optical image. Red dashed lines correspond to the edges of the monolayer graphene and the blue dashed line shows the thin hBN on top. Black dashed line shows the contacting graphite. (b) AFM images. (c) Raman maps.

E4: Twist-angle dependent QTM measurements

Graphite/graphite conductance measurements: for the first proof-of-principle QTM measurements, we have performed out-of-plane conductance measurements $dI/dV$ between a graphite coated tip and a graphite coated substrate, which were performed while continuously varying the twist angle $\theta$ in direct van der Waals (vdW) contact. We find the measured conductance traces to be highly reproducible across multiple rotational cycles, underscoring the mechanical stability and robustness of the quantum twistable mechanical QTM junction throughout the experiment (see Fig. 11). The differential conductance displays a distinct mirror symmetry centered around $\theta = 30°$, where it reaches a pronounced minimum. As the twist angle decreases from this point toward $\theta = 0°$, the conductance increases steadily. However, at very small twist angles, this growth saturates into a plateau, primarily constrained by the finite resistance in regions distant from the junction apex—commonly referred to as contact resistance. The conductance profile also reveals sharp peaks at specific angles, namely $\theta = 21.8°$ and $\theta = 38.2°$. These angles correspond to commensurate stackings with small unit cells, where the atomic lattices of the two monolayers align periodically in real space, giving rise to enhanced electronic coupling. This behavior is consistent with previously observed features in similar van der Waals interfaces performed in the original QTM experiments [1].

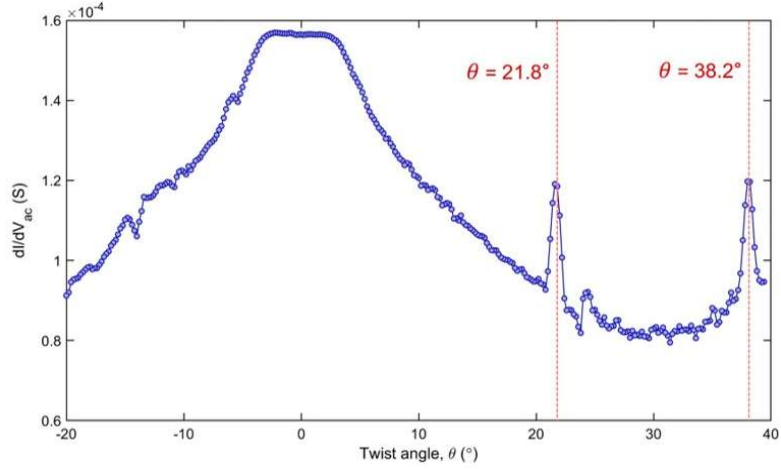

Fig. 11: Measured conductance $dI/dV$ vs. rotation angle $\theta$, between two graphite layers showing peaks at the commensurate configurations with the smallest moiré unit cells.

MLG/hBN/MLG tunneling measurements: the sample is biased with both D.C. and A.C. voltages to simultaneously measure the tunneling current $I$ and differential conductance $dI/dV$ at the tip. Representative measurements are shown in Fig. 12a and b, which show the typical nesting and onset features, which are consistent with the original QTM measurements [1]. In addition, our data shows the nonlinear onset and split nesting features, that are discussed in greater detail in the main part. In order to reveal even finer spectral features, we take a numerical derivative $d^2I/dV^2$ in Fig. 12c.

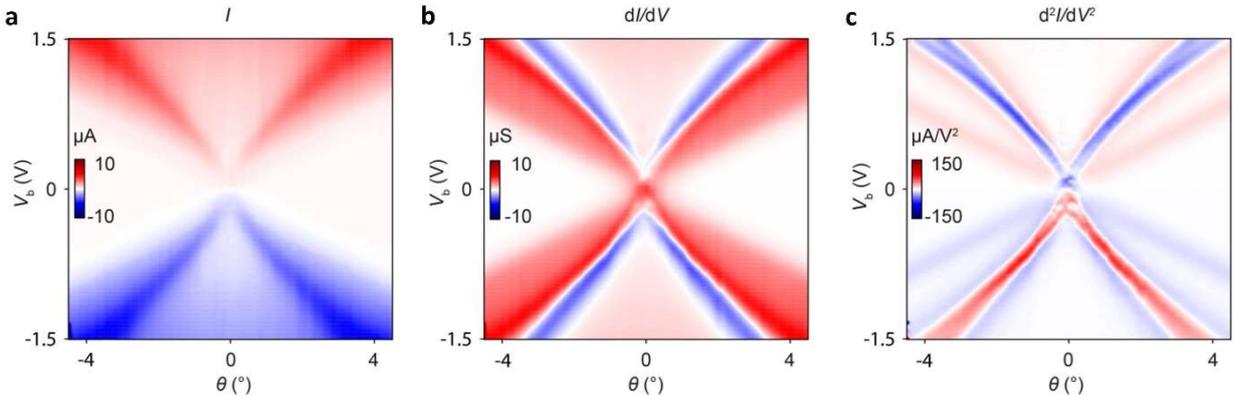

Fig. 12: Elastic tunneling spectroscopy data as a function of twist-angle $\theta$ and bias voltage $V_b$. (a) Measured D.C. tunneling current $I$. (b) Measured differential conductance $dI/dV$. (c) Numerical derivative of the differential conductance $d^2I/dV^2$.

# E5: Multitude of MLG/hBN/MLG tunneling spectra

We have conducted extensive MLG/hBN/MLG tunneling spectroscopy measurements on two separate devices, employing three different tips and probing multiple distinct regions within each device. In total, this yielded nine independent datasets, allowing us to systematically investigate the robustness and reproducibility of our observations. Across all these measurements, we consistently detected the splitting of the nesting condition, a feature that emerges due to the nonlinear dispersion of the band structure. This splitting suggests that the interaction effects play a significant role in shaping the electronic properties of the system. The fact that we observed this behavior across multiple devices, tips, and measurement regions strongly supports the intrinsic nature, and rules out a device dependent origin, like strain or disorder etc.

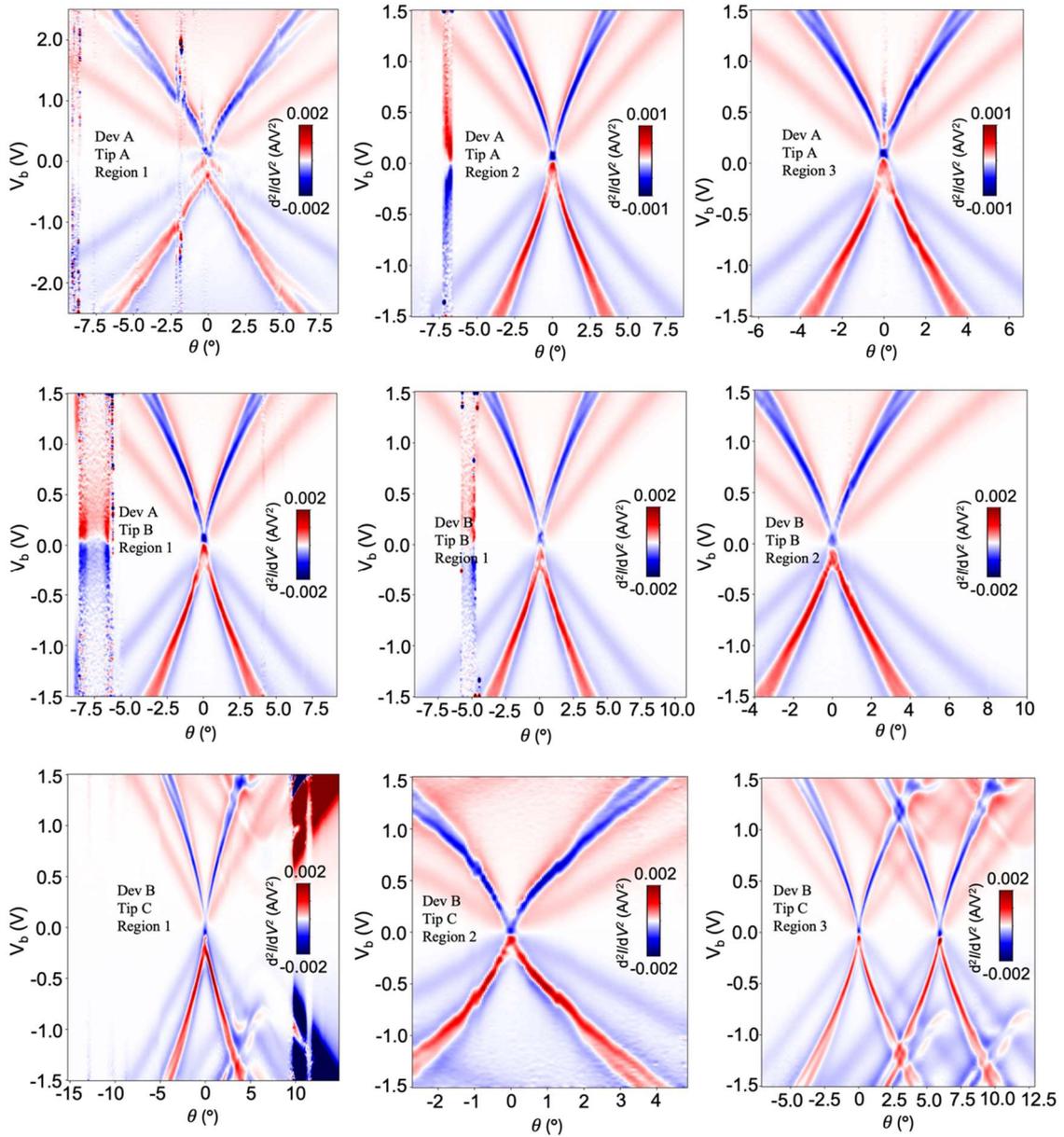

Fig 13: 9 different MLG/hBN/MLG tunneling spectroscopy datasets, showing high reproducibility of the main onset and nesting features in the $d^2I/dV^2$ vs, $\theta$ and $V_b$.

It is worth to note that our device structure incorporates hBN as the tunnel barrier, which has a wide band gap of approximately 6eV. The selection of hBN as the tunnel barrier enables for a high tunneling resolution and allows to apply much higher bias voltages across the sample, as compared to the WSe$_2$ tunnel barriers, that were used in the original QTM measurements [1]. While the breakdown voltage of the WSe$_2$ lies at about 0.8V, we can apply reliably up to $V_b = \pm 2.5$V across the hBN dielectrics, without risking a breakdown. Fig. 13 shows representative MLG/hBN/MLG tunneling measurements up to a $V_b = \pm 2.5$V.

We also want to clarify that two of these data sets (Fig. 13, bottom row) were obtained using Tip C, and exhibited repetitive features within a narrow range of twist angles. We have checked this behavior extensively and believe this occurred because the tip was not sufficiently sharp, likely causing multiple points of the tip to come into contact with the sample during the rotational measurement. We verified and corrected this issue, by repeated the measurements on the same device using a different, sharper tip, which successfully removed the observed anomalies.

## THEORY

Here, we summarize the calculations performed in the main text. A full derivation of the interacting theory of the QTM tunneling spectrum will be given in a future work. In this work, we implement interactions in the Hartree-Fock (HF) approximation. We consider a further rigid doping approximation to remove the well-known HF artifact of a divergent Fermi velocity created by long-range interactions in systems with a Fermi surface. In this approximation, we obtain band structures accounting for the electrostatic effects and the Fock exchange effects (computed at charge neutrality) for all bias voltages. These energies and eigenstates are used as input for the tunneling current formula.

### T1: Single-particle Hamiltonians

We define the lattice vectors and orbital sites

$$\mathbf{a}_1 = a(\frac{1}{2}, \frac{\sqrt{3}}{2})^\mathsf{T}, \qquad \mathbf{a}_2 = a(-\frac{1}{2}, \frac{\sqrt{3}}{2})^\mathsf{T}, \qquad \mathbf{r}_1 = \frac{\mathbf{a}_1 - 2\mathbf{a}_2}{3}, \qquad \mathbf{r}_2 = C_6 \mathbf{r}_1, \qquad \boldsymbol{\tau}_1 = 0, \qquad \boldsymbol{\tau}_2 = \mathbf{r}_2 - \mathbf{r}_1,$$

of the graphene lattice. The K points are $\pm(4\pi/3a, 0)$ where a = 0.246 nm is graphene lattice constant. The Hamiltonian of graphene with nearest neighbor (NN) hopping is

$$h^1(\mathbf{k}) = \begin{pmatrix} 0 & f^*(\mathbf{k}) \\ f(\mathbf{k}) & 0 \end{pmatrix}, \qquad f(\mathbf{k}) = \sum_{j=0}^{2} t(\boldsymbol{\delta}_j) e^{-i\mathbf{k}\cdot\boldsymbol{\delta}_j}, \qquad \boldsymbol{\delta}_j = C_{3z}^j(\mathbf{r}_2 - \mathbf{r}_1).$$

where $t(\delta_j) = -t_0$ and the bare value of the hopping is $t_0 = 2.54$eV. Expanding $f(K + k) = \frac{\sqrt{3}}{2}t_0 a(k_x + ik_y)$ which yields the Dirac Hamiltonian $h(K + k) = v_0 \mathbf{k} \cdot \boldsymbol{\sigma}$ where $v_0 = \frac{\sqrt{3}}{2}t_0 a$ in units where $\hbar = 1$.

We now introduce uniform strain in terms of the symmetric deformation matrix $\mathcal{E}$ in terms of which the new lattice vectors are

$$\tilde{\mathbf{a}}_i = (1 + \mathcal{E})\mathbf{a}_i, \qquad \tilde{\mathbf{K}} = (1 - \mathcal{E}^\mathsf{T})\mathbf{K}.$$

In addition to this geometric effect which shifts the Dirac point, strain modifies the distances between atoms, which leads to differences in the nearest neighbor hopping matrix elements along different directions. The strained Hamiltonian is

$$h^1(\mathbf{k}) = \begin{pmatrix} 0 & \tilde{f}^*(\mathbf{k}) \\ \tilde{f}(\mathbf{k}) & 0 \end{pmatrix}, \qquad \tilde{f}(\mathbf{k}) = \sum_{j=0}^{2} t(\tilde{\boldsymbol{\delta}}_j) e^{-i\mathbf{k}\cdot\tilde{\boldsymbol{\delta}}_j}, \qquad \tilde{\boldsymbol{\delta}}_j = (1 + \mathcal{E})\boldsymbol{\delta}_j$$

We can expand hopping function to first order to obtain

$$t(\tilde{\boldsymbol{\delta}}_j) = t + \nabla t \cdot \mathcal{E}\boldsymbol{\delta}_j = t(1 - \beta \hat{\boldsymbol{\delta}}_j \cdot \mathcal{E}\hat{\boldsymbol{\delta}}_j), \qquad \beta = -\frac{d\log t(r)}{d\log(r)}\bigg|_{r=|\delta_j|}, \qquad \hat{\boldsymbol{\delta}}_j = \frac{\boldsymbol{\delta}_j}{|\boldsymbol{\delta}_j|}$$

where β ≈ 3.14 is a dimensionless parameter characterizing the tight-binding hopping function. Uniaxial strain of strength ε applied along the direction $R(\varphi)\hat{x}$ corresponds to a deformation matrix

$$\mathcal{E} = R(-\varphi)\begin{pmatrix} \epsilon & 0 \\ 0 & -\nu\epsilon \end{pmatrix} R(\varphi)$$

where $\nu = 0.16$ is the Poisson ratio of graphene. Biaxial (isotropic) strain corresponds to $E = \varepsilon 1$. Finally, we introduce the NNN (next nearest neighbor) hopping $t'$ in the Hamiltonian, which gives

$$h^1(\mathbf{k}) = \begin{pmatrix} g(\mathbf{k}) & f^*(\mathbf{k}) \\ f(\mathbf{k}) & g(\mathbf{k}) \end{pmatrix}$$

where $g(\mathbf{k}) = t'\sum_{i=1}^{6} e^{-i\mathbf{k}\cdot C_6^{i-1} a_1}$. Expanding $g(\mathbf{k})$ around the K-point gives $g(\mathbf{k}) = -3t' + \frac{3}{4}at'k^2 + \cdots$. We will drop the constant $-3t'$ as a choice of chemical potential, and take the accepted $t' = 0.3$eV which is consistent with our ab initio calculations.

## T2: Electrostatics and tunneling

We briefly recap the semi-classical formalism used in [1] to model the QTM. These results are physically transparent and accurately describe many features of the device and measurement. A formal derivation within Hartree-Fock theory (which clarifies the novel treatment of Fock exchange, to be discussed in the next section) will be given in a separate paper [11].

In the semi-classical picture, the system is composed of the tip and sample and is described by two band structures which are electrostatically coupled. Denote the Fermi levels of the tip and sample by $\mu_t$ and $\mu_s$ respectively. Under the bias voltage $V_b$, equal and opposite charge accumulates on the tip and sample forming a capacitor with potential energy difference φ. Note that $|\varphi| < |eV_b|$: the Fermi levels in the tip and sample are changed by $V_b$ when charge is accumulated on either layer, which costs kinetic energy. The energy balance is modeled via Kirchoff's law and Gauss's law [1]

$$eV_b = \mu_s - \mu_t + \phi, \qquad \frac{\phi}{d} = \frac{e^2 n_t}{\epsilon_\perp} = -\frac{e^2 n_s}{\epsilon_\perp}$$

where $n_t, n_s$ are the number densities of electrons in the tip and sample measured with respect to charge neutrality, and $d$ is the thickness of the dielectric, which is the distance between the graphene flakes. The parameter $\epsilon_\perp/d$ is the junction capacitance (per area) which is device-dependent and must be fit. We extract $\epsilon_\perp/d \approx 2.4\epsilon_0/\text{nm}$ when we fit the tunneling current, which we now discuss.

The tunneling through the junction can be modeled according to [1] with the Bistritzer-Macdonald tunneling amplitude and linear response theory:

$$I(V_b, \theta) = \frac{8\pi w^2 e}{\hbar} \sum_{kmn,j} |T_j^{mn}(k)|^2 \delta_\gamma\bigl(E_n(k-q_j) - E_m(k) - \phi\bigr)\Bigl(f_t\bigl(E_m(k)\bigr)$$
$$- f_s\bigl(E_n(k-q_j)\bigr)\Bigr),$$

$$T_j^{mn}(k) = U_m^\dagger(k) T_j U_n(k-q_j), \quad T_j = \sigma_0 + \cos\frac{2\pi}{3}\sigma_1 + \sin\frac{2\pi}{3}\sigma_2$$

Here $f_{t,s}(E) = \frac{1}{1+e^{\beta(E-\mu_{t,s})}}$ is the Fermi function, w is the effective tunneling matrix element through the hBN, and $\delta_\gamma(x) = \frac{\gamma/\pi}{x^2+\gamma^2}$ is a smoothed delta function with inverse lifetime $\gamma = 20\text{meV}$. In TBG, $w = w_1 = 110\text{meV}$ describes the direct hopping between the graphene layers. In our experiment, the tunneling barrier strongly suppresses this value. In general, w depends device-specific properties of the barrier, but since it is only a pre-factor in the tunneling current formula above, we do not rely on its exact value. We estimate $w = w_1^2/\Delta_{hBN} \sim 10\text{meV}$ since an electron hop between the tip and sample via a second order process through the thin hBN. Lastly, $E_m(k), U_m(k)$ are the graphene eigen-energies and eigenvectors with $m = \pm$, and

$$q_j(\theta) = [1 - (1+\mathcal{E})R_{-\theta}]C_3^j K$$

are the scattering vectors between the sample K point and the strained/rotated tip K point. In our expression for $I(V_b, \theta)$, we have accounted for spin-valley degeneracy. We have also assumed that the lifetime of either the tip or the sample is very long. This assumption reduces the calculation of the current to a single k-sum, making it fast enough to use for multi-dimensional fitting. We now introduce the dispersion and the interacting parameters.

## T3: Fock-Renormalized Dispersion

We now discuss the effect of interactions within the continuum Dirac approximation with bare velocity $v_0 = 0.85 \times 10^6$ m/s computed from ab-initio calculations in [9]. The Hartree-Fock band structure at charge neutrality can be obtained analytically (as first found in [10]) which allows for us to easily compute the tunneling current map and fit parameters to the experimental data. It can be shown (see the accompanying paper [11]) that the Hartree-Fock bands have the same eigenvectors as the non-interacting Dirac model but with dispersion

$$E(k) = \pm v_0 k \left(1 + \frac{\alpha}{4} \log \frac{4\Lambda/\sqrt{e}}{v_0 k}\right), \quad \alpha = \frac{e^2}{4\pi \epsilon_\parallel v_0}$$

where we have kept the highest order terms in the momentum space cutoff $\Lambda/v_0$. The appearance of the cutoff $\Lambda$ in the dispersion agrees with early renormalization group calculations. Since $\Lambda$ indicates the energy scale of the breakdown of the Dirac model, its value dependence of the precise details of the single-particle hoppings and the effective interaction on the $p_z$ orbitals. In this work, we treat $\Lambda$ as a fitting parameter. The value $\Lambda \approx 3300$ meV we obtain is in good agreement with our microscopic calculations (see the accompanying paper [11]). Finally, $\alpha$ is the graphene fine structure constant which depends on the effective in-plane dielectric constant $\epsilon_\parallel$. We treat $\alpha$ as a fitting parameter and obtain a value of $\alpha = 0.32$ in good agreement with earlier experiments on hBN.

Fig. 14 (left) shows the tunneling spectrum obtained with the interacting dispersion, which shows 3 distinct features. Nonzero tunneling current begins above a critical line on the $\theta, V_b$ plane which we call the onset line (see [1]). There are then two lines where the current attains a local maximum. We refer to these lines as Nesting I and Nesting II. In the linear approximation of the Dirac cone, only one nesting line occurs. We will explain the splitting of the nesting lines and given algebraic expressions for the onset and nesting lines in the next section in the limit of zero temperature and infinite lifetime: $T, \gamma \to 0$.

To conclude this section, we discuss our fitting procedure. There are three free parameters which we use to fit the experimental data: $\epsilon_\perp/d$, $\alpha$, and $\Lambda$. Our approach is to match the onset and nesting I and II features between theory and experiment. These three features constrain the three parameters without overfitting as we now explain. First, the onset line is sensitive to $\alpha$ which renormalizes $v_0 \to v_F$ due to interactions, but is not so sensitive to $\epsilon_\perp/d$, $\Lambda$. This is because the onset line in the Dirac limit is $V_b = v_F K \theta$ (see Ref. 1, and as we derive below). Second, the nesting I line is sensitive to $\epsilon_\perp/d$ (and $v_F$) since in the Dirac limit it reduces to $\phi = v_F K \theta$ and thus is sensitive to the electrostatics determined by $\epsilon_\perp/d$ that convert $V_b$ to $\phi$. Thirdly, the nesting II line is sensitive to $\Lambda$ which produces nonlinearity in the band structure and leads to the split feature. This we explain analytically in the following section T4. In summary, each parameter is associated with a separate feature, but we emphasize that all three parameters are fit simultaneously.

We obtain the approximate onset and nesting lines by extracting the peaks of the $d^2I/dV_b^2$ data (we use the second derivative since the features are clearer on top of the experimental background, as discussed in the main text). In the same manner, we extract the peaks in the second derivative from the theoretically simulated tunneling current maps. Then we optimize $\epsilon_\perp/d$, $\alpha$, and $\Lambda$ via gradient descent. We use the full numerical calculation to account for the effects of temperature, which significantly broadens the onset feature (see Fig. 15) but leaves the nesting features essentially unchanged. This can be expected since the onset feature is due to Pauli blocking set by the Fermi-Dirac distribution, whereas the nesting conditions correspond to peaks in the joint density of states. The best-fit parameters are obtained in the Main Text. Note that we fix the bare value of $\hbar v_0 = 542.1$ meV nm from ab initio calculations and do *not* include it as a fit parameter. Finally, the magnitude of the current is not considered, since it is sensitive to the device area and disorder.

T4: Generalized Onset and Nesting conditions for Monolayer Graphene

We now derive the general forms of the nesting and onset conditions given an isotropic dispersion $\pm E(k)$. The final results are shown in Fig.14 (left - dashed lines) and apply to the usual Dirac Hamiltonian as well as the Fock-interacting dispersion The conditions are derived from the tunneling current at $\gamma = T = 0$ and by neglecting the eigenstate dependence: only the dispersion of the bands in k-space enters. The relative magnitude of the tunneling current will depend on the eigenvector factor (which appear in the numerator and are bounded), but not the positions of the features.

Recall from T2 that at all bias voltages, the electrostatic equilibrium equations are

$$eV_b = \mu_s - \mu_t + \phi, \qquad \frac{\phi}{d} = \frac{e^2 n_t}{\epsilon_\perp} = -\frac{e^2 n_s}{\epsilon_\perp}.$$

Since $n(\mu) = 4 \times \pi k_F^2/(2\pi^2)$, we can solve the second equation via $\mu = E\left(\sqrt{\pi\epsilon\phi/e^2 d}\right)$ which determines, using the first equation, $V_b(\phi)$.

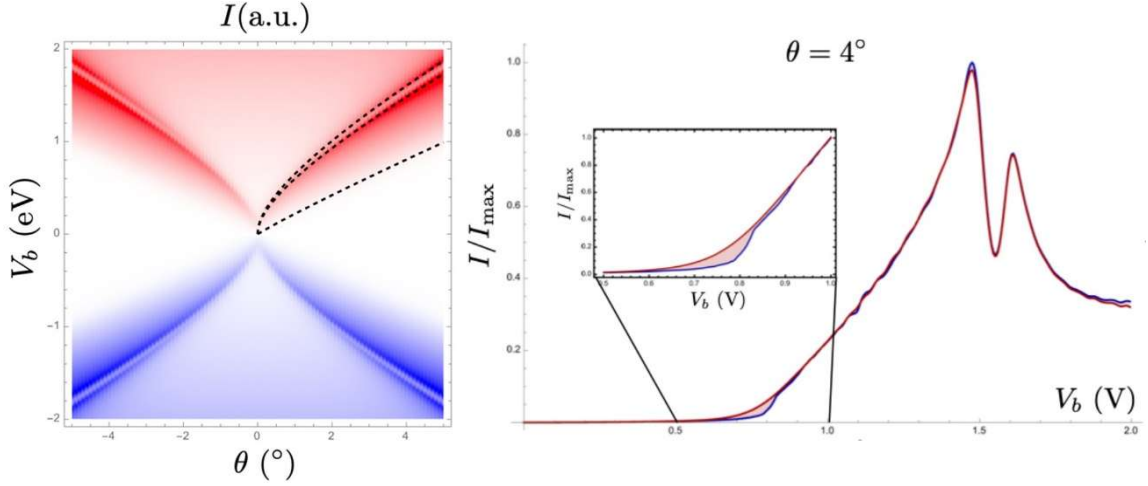

Fig. 14: (left) Tunneling current with onset, nesting I, and nesting II lines dashed. (right) Constant-angle linecut at room temperature (red) and 5K (blue). The two nesting peaks are essentially temperature independent, whereas the onset line is significantly broadened: at room temperature 25K current starts at ~100mV smaller bias voltage than at 5K.

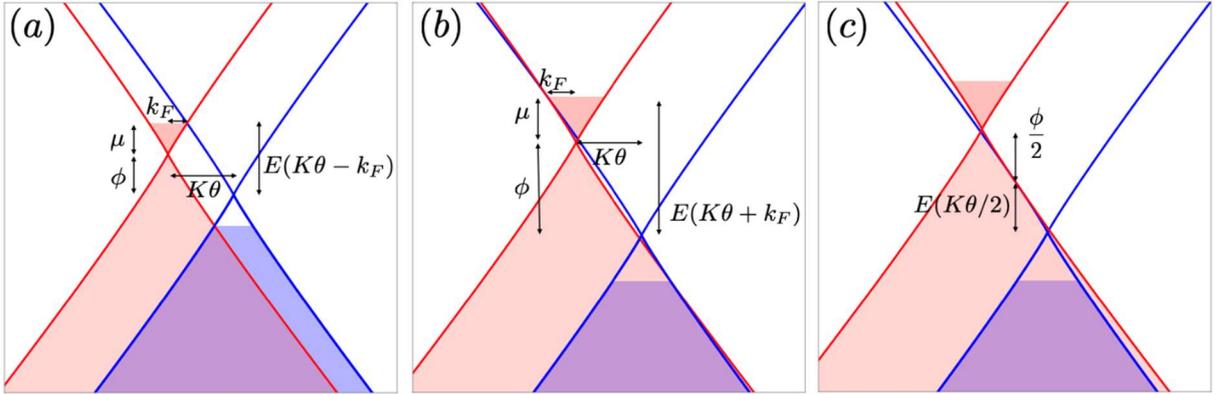

Fig. 15: Band structure diagrams for (a) onset, (b) nesting I, and (c) nesting II. Blue lines are $E_n(k - q_0)$ and red lines are $E_m(k) + \phi$, with Fermi levels shown by shading. $E_m(k)$ is the Hartree-Fock dispersion.

We start by deriving the generalized onset condition. We imagine starting at a large twist angle $\theta$ and small bias so that the Fermi surfaces of the tip and sample do not overlap in k. Then there is no tunneling (at $\gamma = T = 0$). As the bias is increase at fixed angle, the Fermi surfaces will expand until the occupied states in one layer overlap with an unoccupied state in another layer (Fig. 15a). Note that the velocities of the occupied and unoccupied states have opposite sign at this point. Small but nonzero tunneling will start, so this critical bias $V_b(\theta)$ forms the onset line. This equation is

$$\mu = E(K\theta - k_F(\mu)) - \phi$$

where $k_F = E^{-1}(\mu)$ and hence $k_F(\mu) = k_F\left(E(\sqrt{\pi\epsilon_\perp \phi/e^2 d})\right) = \sqrt{\pi\epsilon_\perp \phi/e^2 d}$. In addition to this equation, a closed algebraic system for $V_b(\theta)$ is obtained from the electrostatic conditions (which are always obeyed, even off the onset line). To understand this expression, we take the Dirac example $E(k) = v_F k$ for which the onset equation reads $v_F K\theta - \mu = \phi + \mu$. Using $V_b = \phi + 2\mu$, we obtain $V_b(\theta) = v_F K\theta$ as the onset line. Deviations from the linear Dirac dispersion result in in the $V_b$ vs $\theta$ line describing the onset condition, as observed in experiment in the Main Text.

Next, we study the nesting conditions. As the bias is further increased, the bands will come closer together and tunneling will increase, reaching a maximum when the unoccupied states in one layer overlap the Fermi level and now have *the same sign*, so there will be strong tunneling (see Fig. 15b). This nesting I condition is

$$\mu = E(K\theta + k_F(\mu)) - \phi$$

which again is supplemented with the electrostatic equations to yield a closed set. To understand the solution, we take the Dirac example so that the nesting I equation reads $v_F K\theta + \mu = \phi + \mu$ and thus reduces to $\phi = v_F K\theta$. This can be solved in terms of $V_b$ using the quadratic formula.

Finally, there is a second nesting condition as $V_b$ increases further and unoccupied/occupied states cross and the same Fermi velocity become locally tangent (Fig. 15c). This condition is

$$-E(-K\theta/2) - \frac{\phi}{2} = -E(K\theta/2) + \frac{\phi}{2} \quad \Longrightarrow \quad \phi = 2E(K\theta/2)$$

which is simple enough to yield a closed form solution using the electrostatic equations. We find the nesting II condition is

$$V_b = 2E\left(\frac{\sqrt{\pi\epsilon\, 2E(K\theta/2)}}{e^2 d}\right) + 2E(K\theta/2)$$

which is identical to nesting I if the band in the case of a Dirac band with constant velocity. Otherwise, the two nesting differ.

## T5: Numerical investigation of parameters

We now numerically investigate the effect of various parameters on the tunneling maps to qualitatively understand their effects in interpreting the data. As discussed in the main text, we have theoretically examined the origin of the non-linear dispersion and its impact on the QTM spectra. The most basic model for graphene's band structure is the single-particle nearest-neighbor (NN) hopping model, which at higher energies develops a dispersion that significantly deviates from linearity. Notably, this deviation becomes prominent near the M points, around $\mu \sim 1.5$ eV, where the Fermi surface also exhibits trigonal warping. In Fig. 16 we show theoretical QTM spectrum up to very high bias voltages. One can clearly see that a splitting of the nesting condition due to appears only above $V_b > \pm 5V$, way beyond the experimentally accessible voltage range and experimentally seen splitting at $V_b > \pm 1V$.

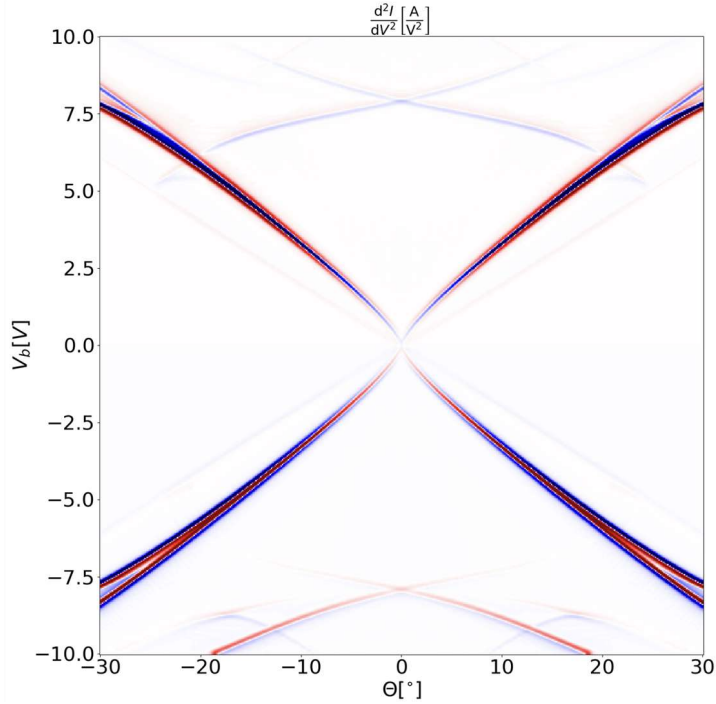

Fig. 16: Non-interacting band structure with only NN $t = 3.2$eV calculated for a large range of bias voltage. Note that this calculation does not consider the dielectric breakdown of the hBN, which will obscure the large bias regions of the plot in a real experiment.

T5.1: Effect of Junction Capacitance

We study the effect of the areal junction capacitance $\epsilon_\perp/d$ on the QTM spectrum, which is shown in Fig. 17. We observe that the onset line is nearly unaffected, while the bending of the nesting line is altered due to the electrostatics. The value of $\epsilon_\perp/d$ is device-dependent. We find a typical value of $\epsilon_\perp/d \approx 2.4\epsilon_0/nm$ in Device A. Since the out-of-plane hBN dielectric constant is $\sim 3\epsilon_0$ and the interlayer distance is $\sim 1nm$, this agrees at the order-of-magnitude level. Detailed modeling of the graphene-monolayer hBN-graphene interface is beyond the scope of this work, and we simply use $\epsilon_\perp/d$ as a fit parameter.

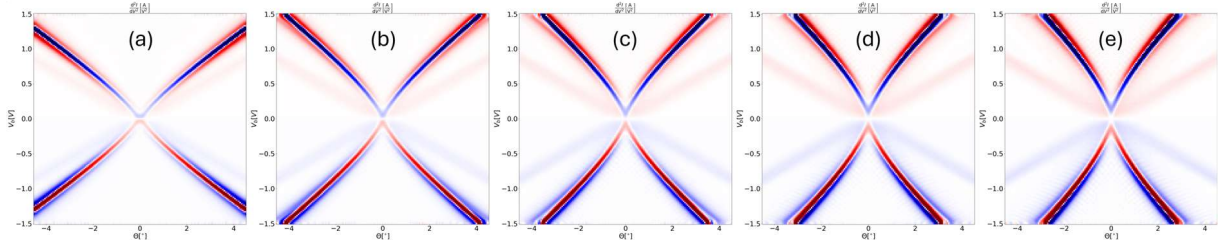

Fig. 17: Non-interacting ($t = 3.2eV$) calculation of $d^2I/dV_b^2$ calculated for different values of $\epsilon_\perp/d = 0.5, 1.5, 2.5, 3.5, 4.5\ \epsilon_0$/nm, the areal junction capacitance.

T5.2: Effect of strain

We also investigate the effect of strain by studying two types: uniaxial heterostrain, and isotropic expansion/contraction, i.e. biaxial strain. Fig. 18 shows that uniaxial strain splits the nesting lines into 3 branches at all values of the bias, since the 3-fold symmetry of the K points is broken. On the other hand, isotropic biaxial strain has the effect of opening a "gap" in the QTM spectrum at zero bias, since the Fermi surfaces of the two layers are shifted apart by the change in lattice constant. Neither possibility is a match to experiment.

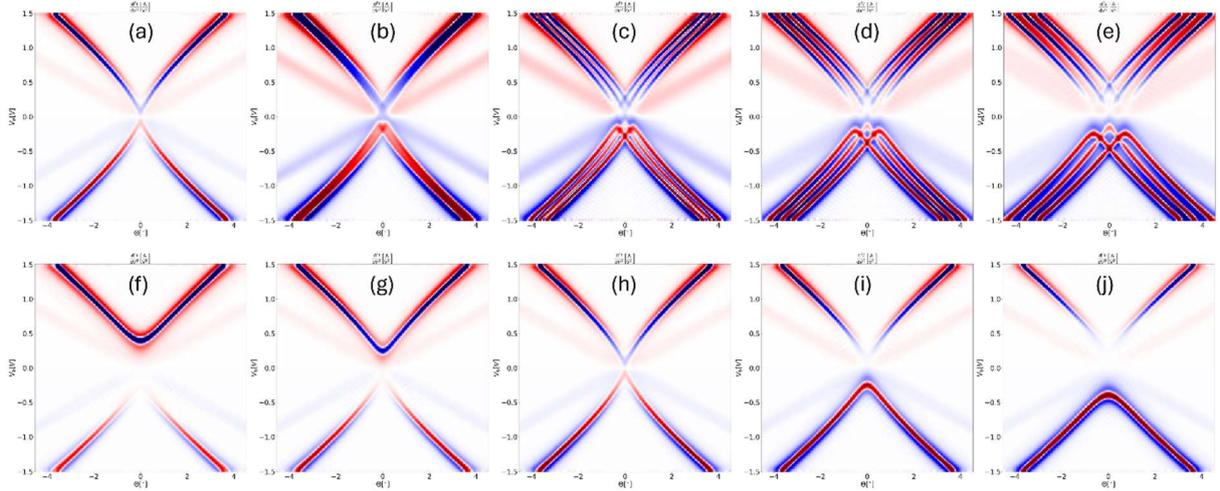

Fig. 18: Non-interacting $d^2I/dV_b^2$ calculated for different values of strain. a-e) correspond to a Poisson ratio of - 0.16 and uniaxial strain of 0.0%, 0.25%, 0.5%, 0.75%, 1.0%, respectively. f)-j) correspond to biaxial strain of -1.0%, -0.5%, 0.0%, 0.5%, 1.0%, respectively. We only include strain in the tip, i.e. we assume the sample is unstrained.

T5.3: Effect of NNN hopping

Another possible source of nonlinearity is the NNN hopping, which is responsible for particle-hole breaking in the Dirac band structure. Ab initio calculations predict $t' = 300$meV in line with experimental reports. This value of the NNN hopping is too small to yield observable splitting within the accessible range of bias voltage in this experiment, although larger values can, as seen in Fig. 19. However, even the qualitative behavior of the splitting induced by unreasonably large NNN hoppings is distinct from the interaction-induced splitting. Notable, the nesting II branch is continuous for NNN hoppings, whereas for interactions the nesting I branch is continuous. The latter case is in agreement with device A and B. In devices where the interaction is suppressed, we predict NNN hopping to cause splitting at $V_b \sim 1.5\text{eV}$ for $\epsilon/d = 2.4\epsilon_0/nm$. Finally, Fig. 19 shows the tunneling spectrum computed over a very large bias and twist angle range.

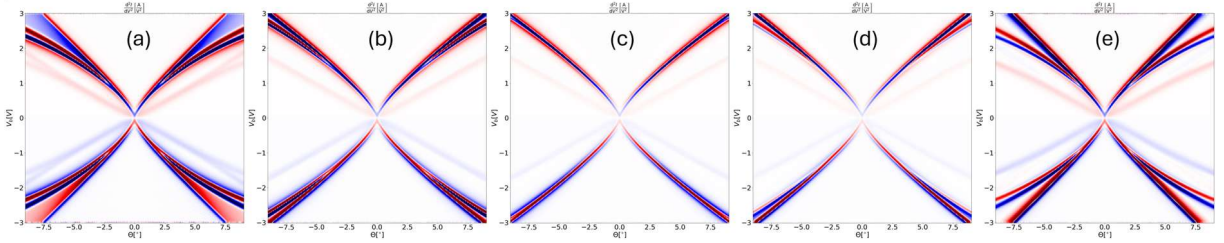

Fig. 19: Non-interacting ($t = 3.2eV$) calculation of $d^2I/dV_b^2$ with NNN hopping calculated for different values of $t'$ = -0.9, -0.3, 0.0, 0.3, 0.9eV.

T5.4: Effect of interaction

We investigate the effect of Coulomb interaction strength on the band structure of monolayer graphene in Fig. 20 by varying α and fixed best-fit parameters $\epsilon_\perp/d, \Lambda$. When the interaction strength vanishes, the nesting lines show no splitting. Turning on $\alpha$ increases the effective $v_F$ and shows splitting within 1.5V at large enough values of $\alpha$. Our best-fit value is $\alpha = 0.24$.

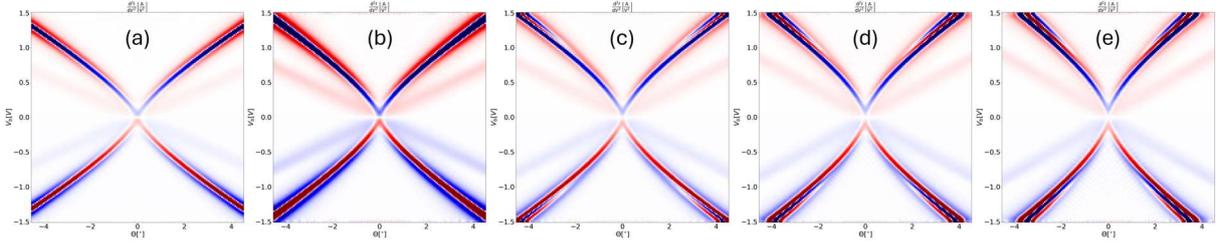

Fig. 20: Band structure calculated for different interaction strength α = 0, 0.1, 0.2, 0.3, 0.4. We see that the strong nesting line in the non-interacting band structure is split into two features and increases the slope of the onset line.

T5.5: Effect of temperature

As discussed, the onset feature is broadened by temperature whereas the nesting peaks are relatively unaffected. In Fig. 21 we calculated the band structure of the monolayer graphene for different temperatures while keeping the interaction term constant and varying the temperature from $T$. We see a broadening in the onset condition as the temperature is increased, whereas the nesting features are largely unchanged since they arise from high density of states.

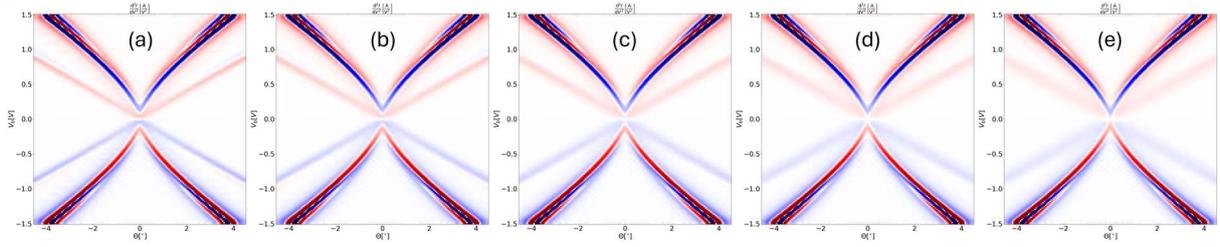

Fig. 21: Interacting band structure of graphene calculated for different temperatures $k_b T = 5, 10, 15, 20, 25$ meV and the best-fit parameters $\alpha, \epsilon_\perp/d, \Lambda$.

T5.6: Effect of lifetime

An additional source of broadening beyond temperatures arises due to the finite electron lifetime $1/\gamma$, which results in a Lorentzian broadening of the spectral functions. This broadening affects the linewidth in $E$-$k$ space but does not influence their peak locations. Fig. 22 illustrates this effect for varying electron lifetimes. Stronger broadening smears out the low-bias Dirac crossing as well as the lowest bias voltage at which splitting of the nesting lines is observed. The lifetime is device-dependent, and can vary from $\gamma = 10$ to $50 meV$.

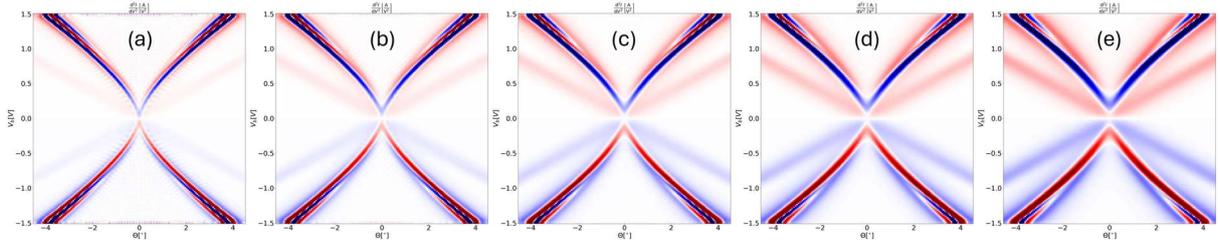

Fig. 22: Interacting $d^2I/dV_b^2$ calculated for $\gamma = 10, 20, 30, 40, 50$ meV respectively and the best-fit parameters $\alpha, \epsilon_\perp/d, \Lambda$.

References:

1. Inbar, A. *et al.* The Quantum Twisting Microscope. *Nature*, 614, (2023).
2. Van Dorp, W. F. *et al.* A critical literature review of focused electron beam induced deposition. *Journal of Applied Physics*, 104(8) (2008).
3. Haddock, D. Characterisation of diamondlike carbon (DLC) laser targets by Raman spectroscopy. *Journal of Physics: Conference Series*, 713, 012007 (2016).
4. Kariman, B. S. *et al.* High dioptric power micro-lenses fabricated by two-photon polymerization. *Optics Express* **32** (27) (2024).
5. Konoshita, K. *et al.* Dry release transfer of graphene and few-layer h-BN by utilizing thermoplasticity of polypropylene carbonate. *NPJ 2D Materials*, 22 (2019).
6. Britnell, L. *et. al.* Electron Tunneling through Ultrathin Boron Nitride Crystalline Barriers. *Nano Lett.* 12, 3, 1707–1710 (2012).
7. Huang, Y. *et. al*. Reliable Exfoliation of Large-Area High-Quality Flakes of Graphene and Other Two-Dimensional Materials. *ACS Nano* 9, 11, 10612–10620 (2015).

8. Shearer, C. *et. al.* Accurate thickness measurement of graphene. *Nanotechnology* 17 (2016).
9. Herzog-Arbeitman, J. *et. al.* Moiré Fractional Chern Insulators II: First-principles Calculations and Continuum Models of Rhombohedral Graphene Superlattices. *Phys. Rev. B* 109, 205122 (2024).
10. Jung. J. *et. al.* Enhancement of nonlocal exchange near isolated band crossings in graphene. *Phys. Rev. B* 84, 085446 (2011).
11. Herzog-Arbeitman, J. *et. al.* Hartree-Fock Band Renormalization in the Quantum Twisting Microscope. *In preparation.*